\documentclass[10pt,journal]{IEEEtran}
\usepackage{xcolor}
\usepackage{cite}
\usepackage{amsmath,amsfonts}
\usepackage{algorithmic}
\usepackage{algorithm}
\usepackage{array}
\usepackage{hyperref}
\usepackage[caption=false,labelfont={footnotesize},textfont={footnotesize}]{subfig}
\usepackage{textcomp}
\usepackage{stfloats}
\usepackage{url}
\usepackage{verbatim}
\usepackage{graphicx}
\usepackage{makecell}
\usepackage{siunitx}
\usepackage{amssymb}
\usepackage[mathscr]{eucal}

\def\BibTeX{{\rm B\kern-.05em{\sc i\kern-.025em b}\kern-.08em
    T\kern-.1667em\lower.7ex\hbox{E}\kern-.125emX}}
\usepackage{balance}
\begin{document}
\title{A 3D Continuous-Space Electromagnetic Channel Model for 6G Tri-Polarized Multi-user Communications}
\author{Yue Yang, Cheng-Xiang Wang, \IEEEmembership{Fellow, IEEE}, Jie Huang, \IEEEmembership{Member, IEEE}, John Thompson, \IEEEmembership{Fellow, IEEE},\\ and H. Vincent Poor, \IEEEmembership{Life Fellow, IEEE}\vspace{-2.5em}
\thanks{
	Manuscript received 12 October 2023; revised 2 April 2024; accepted 17 August 2024. This work was supported by the National Natural Science Foundation of China (NSFC) under Grants 62394290 and 62394291, the Fundamental Research Funds for the Central Universities under Grant 2242022k60006, the Research Fund of National Mobile Communications Research Laboratory, Southeast University, under Grant 2024A05, the U.S National Science Foundation under Grants CNS-2128448 and ECCS-2335876, and EPSRC projects EP/X04047X/1 and EP/Y037243/1. \emph{(Corresponding Authors: Cheng-Xiang Wang and Jie Huang.)}
	
	Y. Yang, C.-X. Wang, and J. Huang are with the National Mobile Communications Research Laboratory, School of Information Science and Engineering, Southeast University, Nanjing 210096, China, and also with the Purple Mountain Laboratories, Nanjing 211111, China (email: \{yueyang, chxwang, j\_huang\}@seu.edu.cn).

	J. Thompson is with the Institute for Digital Communications, School of Engineering, University of
    Edinburgh, Edinburgh EH9 3JL, U.K. (email: john.thompson@ed.ac.uk).

   H. Vincent Poor is with the Department of Electrical and Computer Engineering, Princeton University, Princeton, NJ, USA. (email: poor@princeton.edu).
}}

\markboth{IEEE TRANSACTIONS ON WIRELESS COMMUNICATIONS, VOL.~XX, NO.~XX, MONTH 2024}
{Shell \MakeLowercase{\textit{et al.}}: Bare Demo of IEEEtran.cls for IEEE Journals}

\maketitle
\begin{abstract}
It is envisioned that the sixth generation (6G) and beyond 6G (B6G) wireless communication networks will enable global coverage in space, air, ground, and sea. In this case, both base stations and users can be mobile and will tend to move continuously in three-dimensional (3D) space. Therefore, obtaining channel state information (CSI) in 3D continuous-space is crucial for the design and performance evaluation of future 6G and B6G wireless systems. On the other hand, new 6G technologies such as integrated sensing and communications (ISAC) will also require prior knowledge of CSI in 3D continuous-space. In this paper, a 3D continuous-space electromagnetic channel model is proposed for tri-polarized multi-user communications, taking into account scatterers and spherical wavefronts. Scattered fields are calculated using the method of moments (MoM) with high accuracy. Spherical wave functions are utilized to decompose the dyadic Green's functions that connect the transmitted source currents and the received electric fields. Simulation results demonstrate that transmit power, apertures, scatterers, and sample intervals have significant impacts on statistical properties and channel capacities, providing insights into the performance of continuous-space electromagnetic channel models and the design of future wireless systems.  
\end{abstract}


\section{Introduction}
\IEEEPARstart{I}{t} is expected that, in sixth generation (6G) and beyond 6G (B6G) space-air-ground-sea global integrated networks, base stations may move continually in satellite and vehicular communications, which will cause the wireless channels to be continuous in three-dimensional (3D) space\cite{ref_6Gvision,ref2021S,ref20206G,refTVT22}. Additionally, new technologies and antenna architectures envisioned for 6G have attracted more attention, such as integrated sensing and communications (ISAC)\cite{ISAC,ISAC2,ISAC3} and holographic multiple-input multiple-output (MIMO)\cite{HMIMO,HRS,HMIMO-2020,HMIMO-2022}. The emergence of these technologies creates a need to perceive the surrounding physical environment and channel state information (CSI) at any position in the communication environment, which can be obtained from a 3D continuous-space channel model for near-field communications \cite{CSEM-2008,CSEM-2018}. This channel model has motivated the development of electromagnetic information theory (EIT), which integrates electromagnetic theory and information theory \cite{WCM,WCM2,MD08,EIT-2008,EIT2,ref2018MD}.

Channel modeling is the bridge linking electromagnetic theory and information theory, and is the fundamental basis for network planning and optimization, system performance evaluation, deployment, and standardization\cite{CM0,JSAC,CM2,CM3,S-CM}. Channel models are divided into deterministic channel models and stochastic channel models \cite{ref2022F}. Deterministic channel models mainly include ray-tracing channel models and electromagnetic channel models. Stochastic channel models include geometry-based stochastic models (GBSMs), correlation-based stochastic models (CBSMs), and beam-domain channel models (BDCMs). Most stochastic channel models present difficulties in obtaining CSI and ignore the tri-polarization antennas in the near-field region. However, hybrid channel models that combine deterministic and stochastic channel models begun attracting more attention \cite{RG-2018,EM-2007,THz-HC,HCM,HCM1,HCM2,HCM3}. In \cite{RG-2018}, a vehicle-to-vehicle tunnel channel model using a ray-tracing deterministic model and a stochastic graph model was analyzed. In \cite{EM-2007}, an electromagnetic MIMO channel model using the method of moments (MoM) and the uniform theory of diffraction (UTD) was proposed. In \cite{THz-HC,HCM,HCM1}, terahertz channel measurements were carried out and hybrid channel models including ray-tracing deterministic large-scale fading and GBSM small-scale fading were proposed. In \cite{HCM2}, a satellite hybrid channel model was established, where the deterministic model used the Fresnel-Kirchhoff methodology, and the stochastic model adopted the ITU-R model. In \cite{HCM3}, a hybrid millimeter wave channel model using both ray-tracing and propagation graph was adopted and the simulation results were compared with the channel measurement results. In \cite{EM-2017}, numerical methods such as finite-difference time-domain (FDTD), MoM, and finite element method (FEM) solved using Maxwell's equations, were seen to have high accuracy. A conclusion of this prior work is that an electromagnetic channel model is necessary for 6G continuous-space communications.

Tri-polarized antennas have three independent radiation elements, capable of simultaneously radiating independent polarized signals in three orthogonal directions. These are expected to have better performance in spectral efficiency than single polarized antennas by utilizing polarization diversity \cite{dual-pol}. In \cite{nature}, tri-polarized antennas were proposed by using electromagnetic polarization, which showcased that utilizing six distinguishable electrical and magnetic polarization states can lead to a three-fold increase in channel capacity. In \cite{JT_tri}, a 3D electromagnetic vector sensor was considered in different antenna configurations, which signified the feasibility of tri-polarized antennas as transmitter (Tx) and receiver (Rx). In \cite{MIMO_cube}, the compact MIMO antennas in a cube shape showed high performance in rich-scattering environments. In \cite{tri-pol}, the authors demonstrated that tri-polarized antennas can achieve three times the capacity of uni-polarized antennas according to channel measurements. Nevertheless, this paper assumed that all multipath components have the same delay and the Rx is in the far-field of the Tx, which is unrealistic for envisioned 6G communication systems. In \cite{Liwei}, a tri-polarized multi-user holographic MIMO channel model was proposed and a precoding scheme was designed to realize higher capacity. However, this model did not consider the scatterers in the communication environments. Therefore, the study of tri-polarized multi-user communications considering both near-field effects and the existence of scatterers is of interest.

It is expected that, in 6G wireless communications, channels will evolve from a discrete space to a continuous space, and the dyadic Green's function is used to link the continuous currents at the Tx and the electromagnetic fields at the Rx \cite{GFbook}. In \cite{CSEM-2008}, a continuous-space single-input single-output (SISO) electromagnetic channel model established by plane-wave Green's functions was proposed, and it used eigenfunctions to decompose the transmitted currents and the received electric fields. In \cite{CSEM-2018}, the capacity of the continuous-space channel model with limited transceiver aperture was calculated and the effects of physical loss, radiation, and the Q factor of dielectric on capacities were analyzed. In \cite{DOF-2020} and \cite{Green-2022}, the degree of freedom (DoF) of a spatial stationary holographic MIMO channel and a near-field tri-polarized channel using dyadic Green's functions in isotropic scattering environments were studied, respectively. The influences of antenna apertures on the DoF were also analyzed. In \cite{MU-2022}, the analysis of a spatial stationary electromagnetic holographic MIMO channel using a Fourier plane wave series expansion was proposed. However, the Fourier plane wave expansion assumes that electromagnetic waves are a superposition of a series of plane waves, which may be not applicable in the near field. In \cite{EIT-2008} and \cite{ref1999T}, the concepts of spatial DoF and the capacity of electromagnetic channels were proposed, and the characteristics of electromagnetic sources were studied. The information capacity for the given additive white Gaussian noise model was calculated considering source norm constraints, radiated power constraints, and combined source norm and radiated power constraints. In \cite{EIT2}, the number of users that can be served in multi-user MIMO systems from the electromagnetic field viewpoint was studied. In \cite{ref_book}, the channel capacities of multi-user MIMO wireless communications were calculated. In \cite{ref15Fran} and \cite{GF}, two-dimensional (2D) channel models based on the Green's function and Fourier plane wave expansions were analyzed. In general, most works related to the continuous-space electromagnetic channel model have adopted dyadic Green's functions to model the radiation process and decomposed the functions with appropriate eigenfunctions, or have used the singular value decomposition (SVD) method of the radiation operator. However, the existing continuous-space electromagnetic channel models are 2D and spatially stationary, consider the isotropic scattering environment, and utilize the Fourier plane wave series expansion.

To the best of the authors' knowledge, a 3D continuous-space electromagnetic channel model with tri-polarized antennas in multi-user communications considering rich-scattering environments, non-stationarity, and near-field spherical wavefronts is still missing in the literature. To explore the performance of continuous-space communications, this paper proposes a 3D electromagnetic channel model, which is applied to design an iterative algorithm for calculating the optimal Tx current. The accuracy of this channel model is confirmed through comparison with full-wave simulations. It is demonstrated that the capacity of the continuous-space channel can be improved when considering rich-scattering environments and near-field effects. The novelties and contributions of this paper are summarized as follows.

1) A 3D continuous-space electromagnetic channel model for 6G tri-polarized multi-user communications based on EIT is proposed. This channel model uses stochastic dyadic Green's function to represent the radiation process and uses spherical wave functions to decompose the tri-polarized currents and electric fields.

2) Scattered electric fields are calculated by using the MoM, and the non-stationarity and near-field spherical wavefronts are represented by the birth-death process and spherical wave functions, respectively. Additionally, an efficient and accurate iterative algorithm is designed to calculate the optimal current. 

3) Simulations are conducted by using electromagnetic full-wave simulation software, and the influences of scattering, antenna aperture, and sample interval on statistical properties and channel capacities are investigated, which are beneficial to the design of 6G antennas and systems. 

The rest of this paper is organized as follows. In Section~II, a novel 3D continuous-space electromagnetic tri-polarized channel model considering scattering, non-stationarity, and near-field spherical wavefronts for multi-user communications is proposed. In Section~III, statistical properties and channel capacities of the proposed channel model are derived. Then, numerical simulation results and analysis are shown in Section IV. Finally, the conclusions of this paper are drawn in Section~V.

In this paper, the following notation is defined. Symbols in bold font represent vectors, such as $\textbf{r}$, $\textbf{r}'$. Given a complex number, $\{\cdot\}^\dagger$, $\{\cdot\}^\text{T}$, and $\{\cdot\}^\ast$ are its conjugate transpose, transpose, and conjugate, respectively. In addition, $i$ is denoted as the imaginary unit, which follows $i^2=-1$. Unless specified, the vectors in this paper are column vectors.

\section{A 3D Continuous-Space Electromagnetic Tri-polarized Channel Model}
In this section, a 3D continuous-space electromagnetic channel model considering scattering, non-stationarity, and near-field spherical wavefronts is proposed for tri-polarized multi-user communications. Then, the radiation process is modeled by Green's function, which is the bridge connecting the transmitted currents and received electric fields. The non-stationarity is modeled by the birth-death process. Finally, the spherical wave functions are used to decompose currents, electric fields, and dyadic Green's functions.
\subsection{Radiation Process}
\begin{figure*}[tb]
	\centering
	\includegraphics[width=0.4\textheight]{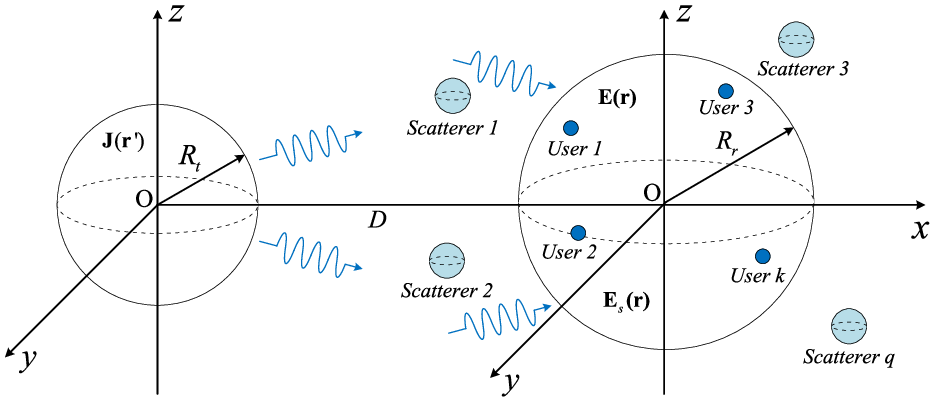}
	\caption{A 3D continuous-space electromagnetic channel model for 6G multi-user communications.}
	\label{Fig1}
\end{figure*}

To construct the continuous-space electromagnetic channel model, the radiation process is first introduced. In Fig.~\ref{Fig1}, a novel 3D electromagnetic channel model for multi-user communications in the continuous space is shown. The current source is distributed in the sphere at the Tx, and users are located in the sphere at the Rx. Scatterers are distributed near the users to reflect the influence of physical communication environments. The distance between the Tx and Rx is represented as $D$, and the radii of the Tx and Rx spheres are $R_t$ and $R_r$, respectively. The volumes of the Tx and Rx spheres are $V_t$ and $V_r$, respectively.

In the free space, the radiation process of the source currents and electric fields can be expressed as \cite{GFbook}
\begin{equation}
    \textbf{E}(\textbf{r})= i\omega \mu \int_{V_t} \overline{\textbf{G}}(\textbf{r},\textbf{r}')\mathbf{J}(\textbf{r}')d^3\textbf{r}'
    \label{equ_1}
\end{equation}
where $\textbf{r}$ is the position at the Rx, $\textbf{r}'$ is the position at the Tx, $\omega$ denotes the angular frequency, $\mu$ is the permeability in free space, $\mathbf{J}(\textbf{r}')$ is the current density at position $\textbf{r}'$, and $\overline{\textbf{G}}(\textbf{r},\textbf{r}')$ is the dyadic Green's function, which is defined as\cite{Class_Green}
\begin{equation}
	\begin{aligned}
		\overline{\textbf{G}}(\textbf{r},\textbf{r}')=&\left(\overline{\textbf{I}}+\frac{\nabla\nabla}{k^2}\right)g(\textbf{r},\textbf{r}')=\left(1+\frac{i}{kr}-\frac{1}{k^2r^2}\right)\overline{\textbf{I}}g(\textbf{r},\textbf{r}')\\&+\left(-1-\frac{3i}{kr}+\frac{3}{k^2r^2}\right)\hat{\textbf{r}}\hat{\textbf{r}}g(\textbf{r},\textbf{r}')
	\end{aligned}
	\label{equ_2}
\end{equation}
where $\overline{\textbf{I}}$ is the unit dyad, $\nabla$ is the differential operator, $k=\omega\sqrt{\mu\varepsilon}$ is the wave number in free space, in which $\varepsilon$ is the permittivity. The distance and unit vector of direction between the Tx and Rx point are denoted as $r=|\textbf{r}-\textbf{r}'|$ and $\hat{\textbf{r}}$, respectively. The scalar Green's function $g(\textbf{r},\textbf{r}')$ in free space can be given by 
\begin{equation}
	g(\textbf{r},\textbf{r}')=\frac{e^{ik|\textbf{r}-\textbf{r}'|}}{4\pi|\textbf{r}-\textbf{r}'|}=\frac{e^{ikr}}{4\pi r}.
	\label{equ_green}
\end{equation}
The Green's function $\overline{\textbf{G}}(\textbf{r},\textbf{r}')$ can be split into three parts according to the terms of $r$, including the $1/r$ term, the $1/r^2$ term, and the $1/r^3$ term, which can represent the far-field, middle-field, and near-field components of the channel, respectively. Therefore, the far-field component is $\overline{\textbf{G}}(\textbf{r},\textbf{r}')|_{F}=\frac{1}{r}(\overline{\textbf{I}}-\hat{\textbf{r}}\hat{\textbf{r}})e^{ikr}$, the middle-field component is $\overline{\textbf{G}}(\textbf{r},\textbf{r}')|_{M}=\frac{i}{kr^2}(\overline{\textbf{I}}-3\hat{\textbf{r}}\hat{\textbf{r}})e^{ikr}$, and the near-field component is $\overline{\textbf{G}}(\textbf{r},\textbf{r}')|_{N}=-\frac{1}{k^2r^3}(\overline{\textbf{I}}-3\hat{\textbf{r}}\hat{\textbf{r}})e^{ikr}$.  

In wireless communications, as the size of antenna arrays increases and the carrier frequency rises, the Rayleigh distance will increase and the near-field effect needs to be considered. Therefore, the near-field components of the radiated field and spherical wavefronts are included in the proposed channel model. To provide a more accurate representation of the near-field effect and connect the radiation process with the communication process, spherical wave functions are used to decompose source currents and dyadic Green's functions. This process is also called the SVD of the radiation operator. The spherical wave functions are defined as \cite{waves}
\begin{equation}
	\textbf{U}_{nml}(r,\theta,\varphi)=\left\{
		\begin{aligned}
			&\nabla\times \textbf{r}h_n^{(1)}(kr)Y_{nm}(\theta,\varphi),&l=1\\
			&\frac{1}{k}\nabla\times\nabla\times \textbf{r}h_n^{(1)}(kr)Y_{nm}(\theta,\varphi),&l=2			
		\end{aligned}
	\right.
	\label{equ_U}
\end{equation}
and 
\begin{equation}
	\textbf{V}_{nml}(r,\theta,\varphi)=\left\{
	\begin{aligned}
		&\nabla\times \textbf{r}j_n^{(1)}(kr)Y_{nm}(\theta,\varphi),&l=1\\
		&\frac{1}{k}\nabla\times\nabla\times \textbf{r}j_n^{(1)}(kr)Y_{nm}(\theta,\varphi),&l=2			
	\end{aligned}
	\right.
	\label{equ_V}
\end{equation}
where $\textbf{r}=r\hat{\textbf{r}}$, $l=1$ and $l=2$ represent the transverse electric (TE) and transverse magnetic (TM) mode, respectively. The subscript $n,m\in \textbf{Z}^3$ are the orders of spherical harmonics, where $n=1,2,3,...$, $-n\le m\le n$. In (\ref{equ_U}), $h_n^{(1)}(x)$ is the $n$th order spherical Hankel function of the first kind and it is defined as $h_n^{(1)}(x)=j_n(x)+iy_n(x)$, where $j_n(x)$ and $y_n(x)$ are the $n$th order spherical Bessel functions of the first kind and the second kind, respectively. They are defined as
\begin{equation}
	j_n(x)=\sqrt{\pi/2x}J_{n+0.5}(x)
\end{equation}
and 
\begin{equation}
	y_n(x)=\sqrt{\pi/2x}Y_{n+0.5}(x)
\end{equation}
where $J_{n}(x)$ and $Y_{n}(x)$ are the $n$th order Bessel function of the first kind and the second kind, respectively. The $n$th order Bessel function of the first kind is calculated as
\begin{equation}
	J_n(x)=\sum_{m=0}^{\infty}\frac{(-1)^m}{m!\Gamma(m+n+1)}\left(\frac{x}{2}\right)^{2m+n}
\end{equation}
where $\Gamma(\cdot)$ is the Gamma function and $m!$ denotes the factorial of $m$. The $n$th order Bessel function of the second kind is  
\begin{equation}
	Y_n(x)=\frac{J_n(x)\cos(n\pi)-J_{-n}(x)}{\sin(n\pi)}.
\end{equation}

In (\ref{equ_U}) and (\ref{equ_V}), the spherical harmonics $Y_{nm}(\theta,\varphi)$ are defined as 
\begin{equation}
	Y_{nm}(\theta,\varphi)=\sqrt{\frac{(n-|m|)!(2n+1)}{(n+|m|)!4\pi}}P_n^{|m|}(\cos\theta)e^{im\varphi}
\end{equation}
where $\theta$ and $\varphi$ are the azimuth and elevation angles, $P_n^{m}(x)$ is the associated Legendre function of the first kind of order $m$ and degree $n$, which is 
\begin{equation}
	P_n^{m}(x)=(1-x^2)^{m/2}\frac{d^mP_n(x)}{dx^m}
\end{equation}
where $m<n$ and $P_n(x)$ is the $n$th order Legendre function of the first kind, which can be calculated as
\begin{equation}
	P_n(x)=\frac{1}{2^nn!}\frac{d^n(x^2-1)^n}{dx^n}.
\end{equation}

The dyadic Green's function can be decomposed according to spherical wave functions $\textbf{U}_{nml}$ and $\textbf{V}_{nml}$ and is expressed as\cite{GF,waves}
\begin{equation}
	\overline{\textbf{G}}(\textbf{r},\textbf{r}')=ik\sum_{n=1}^{\infty}\sum_{m=-n}^{n}\sum_{l=1}^{2}\frac{1}{n(n+1)}\textbf{U}_{nml}(\textbf{r})\textbf{V}_{nml}^\dagger(\textbf{r}')
	\label{equ_de}
\end{equation}
where $|\textbf{r}'|<|\textbf{r}|$. For simplicity, we assume that $p=2(n(n+1)+m-1)+l$ and we can obtain
\begin{equation}
	\sum_{n=1}^{N}\sum_{m=-n}^{n}\sum_{l=1}^{2}\textbf{U}_{nml}=\sum_{p=1}^{2N(N+2)}\textbf{U}_p.
\end{equation}
Therefore, (\ref{equ_de}) can be simplified as
\begin{equation}
	\overline{\textbf{G}}(\textbf{r},\textbf{r}')=ik\sum_{p=1}^{\infty}\frac{1}{n(n+1)}\textbf{U}_{p}(\textbf{r})\textbf{V}^\dagger_{p}(\textbf{r}'), |\textbf{r}'|<|\textbf{r}|.
	\label{equ_G}
\end{equation}

The spherical wave functions have orthogonality within the volumes of the Tx and Rx, so the following formula holds
\begin{equation}
	\int_{V_r}\textbf{U}_{p1}(\textbf{r})\cdot\textbf{U}^\ast_{p2}(\textbf{r})d^3\textbf{r}=0,p1\neq p2 
\end{equation}
\begin{equation}
	\int_{V_t}\textbf{V}_{p1}(\textbf{r})\cdot\textbf{V}^\ast_{p2}(\textbf{r})d^3\textbf{r}=0,p1\neq p2.
\end{equation}
For the convenience of derivation, $\textbf{U}_{p}$ and $\textbf{V}_{p}$ are normalized and can be written as
\begin{equation}
	\textbf{u}_p(\textbf{r})=\frac{\textbf{U}_p(\textbf{r})}{\sqrt{\int_{V_r}|\textbf{U}_p(\textbf{r})|^2d^3\textbf{r}}}
	\label{equ_u}
\end{equation}
\begin{equation}
	\textbf{v}_p(\textbf{r})=\frac{\textbf{V}_p(\textbf{r})}{\sqrt{\int_{V_t}|\textbf{V}_p(\textbf{r})|^2d^3\textbf{r}}}
	\label{equ_v}
\end{equation}
which are mutually orthogonal functions. Therefore, the following formula holds
\begin{equation}
	\int_{V_r}\textbf{u}_{p1}(\textbf{r})\cdot\textbf{u}^\ast_{p2}(\textbf{r})d^3\textbf{r}=\left\{
	\begin{aligned}
		&0,p1\neq p2\\
		&1,p1=p2
	\end{aligned}
	\right.
\end{equation} 
and
\begin{equation}
	\int_{V_t}\textbf{v}_{p1}(\textbf{r})\cdot\textbf{v}^\ast_{p2}(\textbf{r})d^3\textbf{r}=\left\{
	\begin{aligned}
		&0,p1\neq p2\\
		&1,p1=p2
	\end{aligned}
	\right..
\end{equation}                                                              
Then, (\ref{equ_u}) and (\ref{equ_v}) can be substituted into (\ref{equ_G}), and the dyadic Green's function is expressed as
\begin{equation}
	\overline{\textbf{G}}(\textbf{r},\textbf{r}')=ik\sum_{p=1}^{\infty}a_p\textbf{u}_{p}(\textbf{r})\textbf{v}^\dagger_{p}(\textbf{r}'), \textbf{r}'\in V_t, \textbf{r}\in V_r
	\label{equ_G2}
\end{equation}
where 
\begin{equation}
	a_p=\frac{\sqrt{\int_{V_r}|\textbf{U}_p(\textbf{r})|^2d^3\textbf{r}}\sqrt{\int_{V_t}|\textbf{V}_p(\textbf{r}')|^2d^3\textbf{r}'}}{n(n+1)}.
\end{equation}
Similarly, the source current can also be decomposed by spherical wave functions, which is calculated as
\begin{equation}
	\textbf{J}(\textbf{r}')=\sum_{p=1}^{\infty}j_p\textbf{v}_p(\textbf{r}')
	\label{equ_J}
\end{equation}
where $j_p$ can be obtained as
\begin{equation}
	j_p=\int_{V_t}\textbf{J}(\textbf{r}')\cdot \textbf{v}^\ast_p(\textbf{r}')d^3\textbf{r}'.
	\label{j_p}
\end{equation}
Due to the orthogonality and normality of $\textbf{v}_p(\textbf{r}')$, the transmit power is calculated as
\begin{equation}
	\int_{V_t}|\textbf{J}(\textbf{r}')|^2d^3\textbf{r}'=\int_{V_t}\bigg|\sum_{p=1}^{\infty}j_p\textbf{v}_p(\textbf{r}')\bigg|^2d^3\textbf{r}'=\sum_{p=1}^{\infty}|j_p|^2
\end{equation}
which is consistent with the form of power calculation in discrete channels. Then, substituting (\ref{equ_G2}) and (\ref{equ_J}) into (\ref{equ_1}) and we can obtain
\begin{equation}
	\begin{aligned}
		\textbf{E}(\textbf{r})&=-\omega\mu k \int_{V_t} \sum_{p=1}^{\infty} \sum_{p'=1}^{\infty} a_pj_{p'}\textbf{u}_p(\textbf{r})\textbf{v}_p^\dagger(\textbf{r}')\textbf{v}_{p'}(\textbf{r}')d^3\textbf{r}'\\&=-\omega \mu k\sum_{p=1}^{\infty}a_pj_p\textbf{u}_p(\textbf{r})
		\label{equ_p}
	\end{aligned}
\end{equation}
which utilizes the orthogonality and normality of spherical wave functions $\textbf{v}_p$ in volume $V_t$. Assume that $\sigma_p=-\omega \mu k a_p$, electric fields at position $\textbf{r}$ can be written as \cite{ref09Fran}
\begin{equation}
	\textbf{E}(\textbf{r})=\sum_{p=1}^{\infty}\sigma_pj_p\textbf{u}_p(\textbf{r})
	\label{equ_SVD}
\end{equation}
which is the SVD of the radiation process and $\sigma_p\ (p=1,2,...)$ is the singular value of the radiation operator. 
\subsection{Multi-User Communications}
In 6G wireless communications, base stations and users are distributed continuously in space. Assume that there are $K$ users distributed in the volume of the Rx in Fig. \ref{Fig1}. The position of the $k$th user is $\mathbf{r}_k=(R_k,\theta_k,\varphi_k)$. The information sent from the source area to the $k$th user is $s_k$ and the electric field received by the $k$th user is
\begin{equation}
	\textbf{E}(\mathbf{r}_k)=w_{r,k}\textbf{E}_r(\mathbf{r}_k)+w_{\theta,k}\textbf{E}_{\theta}(\mathbf{r}_k)+w_{\varphi,k}\textbf{E}_{\varphi}(\mathbf{r}_k)
	\label{equ_E}
\end{equation}
where $\textbf{E}_r(\mathbf{r}_k)$, $\textbf{E}_{\theta}(\mathbf{r}_k)$, and $\textbf{E}_{\varphi}(\mathbf{r}_k)$ are components of the electric field $\textbf{E}(\mathbf{r}_k)$ at the $k$th user in the directions of $r$, $\theta$, and $\varphi$, respectively. In addition, $w_{r,k}$, $w_{\theta,k}$, and $w_{\varphi,k}$ are gains of the electric field at the $k$th user to the information $s_k$ in the directions of $r$, $\theta$, and $\varphi$, respectively. Substituting (\ref{equ_SVD}) into (\ref{equ_E}) and $\textbf{E}(\mathbf{r}_k)$ can be updated as
\begin{equation}
	\textbf{E}(\mathbf{r}_k)=\sum_{p=1}^{\infty}\sigma_pj_p(w_{r,k}\textbf{u}_{r,p}+w_{\theta,k}\textbf{u}_{\theta,p}+w_{\varphi,k}\textbf{u}_{\varphi,p})
\end{equation} 
where $\textbf{u}_{r,p}$, $\textbf{u}_{\theta,p}$, and $\textbf{u}_{\varphi,p}$ are components of spherical wave functions in the directions of $r$, $\theta$, and $\varphi$, respectively, which are three orthogonal directions in spherical coordinate systems. Therefore, the antennas at the Rx side are tri-polarized.

For reliable communications, the source current $\textbf{J}(\textbf{r}')$ in (\ref{equ_J}) needs to be expanded by choosing suitable weighting factors $j_p$ defined in (\ref{j_p}). Then, user $k$ can retrieve the information $s_k$ intended to be sent by the Tx according to the received electric field $\textbf{E}(\mathbf{r}_k)$, which can be solved by an optimization problem
\begin{equation}
	\begin{aligned}
		\mathcal{P}1: \min_{\boldsymbol{j}}\sum_{k=1}^{K}|\textbf{E}(\mathbf{r}_k)-s_k|^2&=\min_{\boldsymbol{j}}\sum_{k=1}^{K}|\boldsymbol{b}_k^\text{T}\boldsymbol{j}-s_k|^2\\
		s.t. \sum_{p=1}^{\infty}|j_p|^2&\leq P_T
		\label{equ_p2}
	\end{aligned}
\end{equation}
where $P_T$ is the total transmit power and $\boldsymbol{b}_k$ is defined as 
\begin{equation}
	\boldsymbol{b}_k=\left[\begin{array}{c}
		\sigma_{1}(w_{r,k}\textbf{u}_{r,1}+w_{\theta,k}\textbf{u}_{\theta,1}w_{\varphi,k}\textbf{u}_{\varphi,1})\\
		\cdots\\
		\sigma_{p}(w_{r,k}\textbf{u}_{r,p}+w_{\theta,k}\textbf{u}_{\theta,p}w_{\varphi,k}\textbf{u}_{\varphi,p})\end{array}
	\right]
\end{equation}
and the objective function of the optimization problem is to minimize the error between $\textbf{E}(\mathbf{r}_k)$ of all users and the information $s_k$, while satisfying the transmit power constraint. The objective function can be calculated as
\begin{equation}
	\begin{aligned}
		&\sum_{k=1}^{K}|\boldsymbol{b}_k^\text{T}\boldsymbol{j}-s_k|^2=\sum_{k=1}^{K}(\boldsymbol{b}_k^\text{T}\boldsymbol{j}-s_k)(\boldsymbol{j}^\dagger\boldsymbol{b}_k^\ast-s_k^\ast)\\&=\boldsymbol{j}^\dagger\left(\sum_{k=1}^{K}\boldsymbol{b}_k^\ast\boldsymbol{b}_k^\text{T}\right)\boldsymbol{j}-2\text{Re}\left\{\left(\sum_{k=1}^{K}s_k^\ast\boldsymbol{b}_k^\text{T}\right)\boldsymbol{j}\right\}+\sum_{k=1}^{K}|s_k|^2
	\end{aligned}
	\end{equation}
where $\text{Re}\{\cdot\}$ is the real part of complex values, $\sum_{k=1}^{K}\boldsymbol{b}_k^\ast\boldsymbol{b}_k^\text{T}$ is a positive-definite matrix, and the constraint is convex, which can be solved by the Lagrange multiplier and the Karush-Kuhn-Tucker (KKT) conditions. Firstly, the Lagrange function $L(\boldsymbol{j},\lambda), \lambda\geq 0$ is created and written as
\begin{equation}
	\begin{aligned}
		L(\boldsymbol{j},\lambda)=&\boldsymbol{j}^\dagger\left(\sum_{k=1}^{K}\boldsymbol{b}_k^\ast\boldsymbol{b}_k^\text{T}\right)\boldsymbol{j}-2\text{Re}\left\{\left(\sum_{k=1}^{K}s_k^\ast\boldsymbol{b}_k^\text{T}\right)\boldsymbol{j}\right\}\\&+\sum_{k=1}^{K}|s_k|^2+\lambda(\boldsymbol{j}^\dagger\boldsymbol{j}-P_T)\\=&\boldsymbol{j}^\dagger\left(\sum_{k=1}^{K}\boldsymbol{b}_k^\ast\boldsymbol{b}_k^\text{T}+\lambda \textbf{I}\right)\boldsymbol{j}-2\text{Re}\left\{\left(\sum_{k=1}^{K}s_k^\ast\boldsymbol{b}_k^\text{T}\right)\boldsymbol{j}\right\}\\&+\sum_{k=1}^{K}|s_k|^2-\lambda P_T
	\end{aligned}
\end{equation}
where $\textbf{I}$ is an identity matrix and its dimension is the same as the dimension of $\sum_{k=1}^{K}\boldsymbol{b}_k^\ast\boldsymbol{b}_k^\text{T}$. According to convex optimization, the optimal solution $\boldsymbol{j}_{opt}$ needs to follow
\begin{equation}
	\left\{
	\begin{aligned}
		\left.\frac{\partial L(\boldsymbol{j},\lambda_{opt})}{\partial \boldsymbol{j}}\right|\boldsymbol{j}_{opt}&=0\\
		\lambda_{opt}&\geq 0\\
		\boldsymbol{j}_{opt}^\dagger\boldsymbol{j}_{opt}&\leq P_T\\
		\lambda_{opt}(\boldsymbol{j}_{opt}^\dagger\boldsymbol{j}_{opt}-P_T)&=0
	\end{aligned}
	\right.
\end{equation}
and the solution is expressed as
\begin{equation}
	\boldsymbol{j}_{opt}=\left(\sum_{k=1}^{K}\boldsymbol{b}_k^\ast\boldsymbol{b}_k^\text{T}+\lambda_{opt} \textbf{I}\right)^{-1}\left(\sum_{k=1}^{K}s_k\boldsymbol{b}_k^\ast\right)
	\label{equ_36}
\end{equation}
where $\lambda_{opt}$ is an uncertain constant and can be determined by using the bisection method. The computational complexity of calculating (\ref{equ_36}) is $O(K^3)$ due to the matrix inverse operation.

In (\ref{equ_p}), the value of the summation subscript $p$ in the constraint can only be taken to a maximum and value $P$, which is the number of items in the SVD process. In multi-user communications, the value of $P$ can be determined by signal error and transmit power. The signal error is defined as the difference between the received electric fields and the signals transmitted by the source region, and it can be expressed as
\begin{equation}
	err=\frac{\sum_{k=1}^{K}|\textbf{E}(\mathbf{r}_k)-s_k|^2}{\sum_{k=1}^{K}|s_k|^2}.
	\label{err}
\end{equation}
Additionally, the optimal transmit power is defined as
\begin{equation}
	|\boldsymbol{j}_{opt}|^2=\sum_{p=1}^{P}|(j_p)_{opt}|^2.
\end{equation}
Suitable numbers of items in SVD need to minimize the signal transmission error and transmit power. In addition, the value of $P$ cannot be too large, since it will increase the computational complexity in multi-user communication systems.
 
\subsection{Scatterers and Scattered Electric Fields}
In practice, there are various scatterers in communication environments, which have an impact on multi-user communications \cite{ref11RJ}. Assume that scatterers are ideal spherical conductors located on the outside of the Rx sphere. Scatterers may appear and disappear at any time or position, and the generation and recombination of scatterers can be characterized by birth-death processes. The number of rays in the scatterers follows the Poisson distribution, the virtual delay and mean power of each scatterer follow exponential distributions, and the azimuth and elevation angles of scatterers are assumed to be wrapped Gaussian distributions. The positions of the $q$th scatterers can be generated as \cite{CM0}
\begin{equation}
	\mathbf{r}_q(x,y,z) = \frac{e^{\left(-\frac{x^2}{2\sigma^2_{DS}}-\frac{y^2}{2\sigma^2_{AS}}-\frac{z^2}{2\sigma^2_{ES}}\right)}}{(2\pi)^{3/2}\sigma_{DS}\sigma_{AS}\sigma_{ES}}
\end{equation}
where $\sigma^2_{DS}$, $\sigma^2_{AS}$, and $\sigma^2_{ES}$ are variances of Gaussian distributions that describe delay spread, angular spread, and elevation spread of scatterers, respectively. In the time domain, the survival probability $P_\text{sur}(\Delta t)$ of a scatterer in time $\Delta t$ can be computed as
\begin{equation}
	P_\text{sur}(\Delta t)=e^{-\lambda_D\frac{P_f\left(\Delta v^R+\Delta v^T\right)\Delta t}{D_c}}
\end{equation}
where $\lambda_D$ is the recombination rate of scatterers, $P_f$ is the proportion of moving scatterers, $\Delta v^T$ and $\Delta v^R$ are the relative velocities of scatterers at the Tx and Rx region, respectively. The coherence distance $D_c$ is dependent on communication scenarios, and the typical value can be found in \cite{CM3}. In the space domain, the survival probability with sample intervals $\delta_t$ and $\delta_r$ can be calculated as 
\begin{equation}
	P_\text{sur}(\delta_t,\delta_r)=e^{-\lambda_D\frac{\left(\delta_t \cos\left(\beta_E^T\right)+\delta_r \cos\left(\beta_E^R\right)\right)}{D_c}}
\end{equation}
where $\delta_t$ and $\delta_r$ are the sample intervals at the Tx and Rx regions, respectively. The elevation angles of the Tx and Rx sample points are denoted as $\beta_E^T$ and $\beta_E^R$, respectively. 

Therefore, the survival probability of the scatterer within $\Delta t$, $\delta_t$, and $\delta_r$ is written as
\begin{equation}
	P_\text{sur}(\Delta t,\delta_t,\delta_r)=P_\text{sur}(\Delta t)\cdot P_\text{sur}(\delta_t,\delta_r).
\end{equation}
The mean value of newly generated scatterers number $Q_\text{new}$ can be calculated as
\begin{equation}
	\mathbb{E}\{Q_\text{new}\}=\frac{\lambda_B}{\lambda_D}\left(1-P_\text{sur}(\Delta t,\delta_t,\delta_r)\right)
\end{equation}
where $\mathbb{E}\{ \cdot \}$ is the expectation operator, $\lambda_B$ is the generation rate of scatterers.

When there are scatterers, the electric fields received by the user consist of two parts: one is generated by the radiation of the source current $\textbf{J}(\textbf{r}')$, denoted as $\textbf{E}(\textbf{r})$, and the other is the scattered electric field radiated by the induced current on the surface of the ideal conductor scatterers, denoted as $\textbf{E}_s(\textbf{r})$. Assume that the total number of scatterers is $Q$ and the position of the $q$th scatterer is $\mathbf{r}_q=(R_q,\theta_q,\varphi_q)$. In this paper, the MoM is used to calculate the scattered electric fields. Unlike the traditional Born approximation method, the MoM can be used in rich-scattering and 3D communication scenarios. Solving the scattered electric fields is equivalent to solving the induced currents on the surface of the ideal conductor scatterers. Let the surface current density of the $q$th scatterer be $\textbf{J}_q(\textbf{r})$, and it can be expanded using the basis function $\textbf{B}_d(\textbf{r})$, which is written as
\begin{equation}
	\textbf{J}_q(\textbf{r})=\sum_{d=1}^{D}j_{qd}\textbf{B}_d(\textbf{r})
\end{equation}
where $\textbf{B}_d(\textbf{r})=\textbf{V}_{2d-1}(\textbf{r})$ and $j_{qd}$ represents the expansion coefficient of the surface current density of the $q$th scatterer. Electric fields generated by the surface current of the $q$th scatterer can be expressed as
\begin{equation}
	\begin{aligned}
		\textbf{E}_{s,q}(\textbf{r})&=i\omega\mu \int_{S_q}\overline{\textbf{G}}(\textbf{r},\textbf{r}')\textbf{J}_q(\textbf{r}')d^2\textbf{r}'\\&=\sum_{d=1}^{D}j_{qd}\left(i\omega\mu\int_{S_q}\overline{\textbf{G}}(\textbf{r},\textbf{r}')\textbf{B}_d(\textbf{r}')d^2\textbf{r}'\right)
	\end{aligned}
\end{equation} 
where $S_q$ is the surface of the $q$th scatterer. Then, the total scattered electric fields of $Q$ scatterers can be expressed as
\begin{equation}
	\textbf{E}_s(\textbf{r})=\sum_{q=1}^{Q}\textbf{E}_{s,q}(\textbf{r})=\sum_{q=1}^{Q}\sum_{d=1}^{D}j_{qd}\textbf{E}_{s,q}^d(\textbf{r})
	\label{equ_sca}
\end{equation}
where $\textbf{E}_{s,q}^d(\textbf{r})=i\omega\mu\int_{S_q}\overline{\textbf{G}}(\textbf{r},\textbf{r}')\textbf{B}_d(\textbf{r}')d^2\textbf{r}'$. The electromagnetic field boundary condition is utilized to solve the expansion coefficient $j_{qd}$ of the surface current of the scatterer. The boundary condition is that the tangential component of the electric field on the surface of the ideal conductor is $0$. Therefore, the received electric fields are the summation of electric fields generated by the source current $\textbf{J}(\textbf{r})$ and the induced current $\textbf{J}_q(\textbf{r})$. The total electric fields $\textbf{E}_t(\mathbf{r}_k)$ can be calculated as
\begin{equation}
	\begin{aligned}
	\textbf{E}_t(\mathbf{r}_k)=&\sum_{p=1}^{\infty}\sigma_pj_p(w_{r,k}\textbf{u}_{r,p}+w_{\theta,k}\textbf{u}_{\theta,p}+w_{\varphi,k}\textbf{u}_{\varphi,p})\\&+(w_{r,k}\textbf{E}_{r,s}(\mathbf{r}_k)+w_{\theta,k}\textbf{E}_{\theta,s}(\mathbf{r}_k)+w_{\varphi,k}\textbf{E}_{\varphi,s}(\mathbf{r}_k))
	\end{aligned}	
\end{equation} 
where $\textbf{E}_{r,s}(\mathbf{r}_k)$, $\textbf{E}_{\theta,s}(\mathbf{r}_k)$, and $\textbf{E}_{\varphi,s}(\mathbf{r}_k)$ are components of the scattered electric field $\textbf{E}_s(\mathbf{r}_k)$ at the $k$th user in the directions of $r$, $\theta$, and $\varphi$, respectively. 

Assume that there are $N_s$ sample points on the surface of each scatterer, and we can use the tangential boundary condition of electric fields at $Q\times N_s$ points to implement the point matching method in the MoM. The spherical coordinate of the $n$th point selected from the $q$th scatterer is $\mathbf{r}_{qn}=(R_q,\theta_{qn},\varphi_{qn})$ and it follows
\begin{equation}
	\vec{n}\times \left\{\textbf{E}(\mathbf{r}_{qn})+\textbf{E}_s(\mathbf{r}_{qn})\right\}=\textbf{0}
	\label{equ_tang}
\end{equation}
where $\vec{n}$ is the normal vector of the scatterer surface. This expression indicates that the tangential component of the sphere at the field point is zero. Then, substituting (\ref{equ_sca}) into (\ref{equ_tang}) and we can obtain
\begin{equation}
	\sum_{q=1}^{Q}\sum_{d=1}^{D}j_{qd} \{\vec{n}\times\textbf{E}_{s,q}^d(\textbf{r})\}=-\vec{n}\times\textbf{E}(\mathbf{r}_{qn})
\end{equation}
where $\textbf{E}(\mathbf{r}_{qn})=\sum_{p=1}^{\infty}\sigma_pj_p\textbf{u}_p(\mathbf{r}_{qn})$. To obtain an accurate value of $j_{qd}$, the sample points $N_s$ on each scatterer need to be large enough. 

Similarly, multi-user communications considering scattering can be modeled as an optimization problem, which follows
\begin{equation}
	\begin{aligned}
		\mathcal{P}2: \min_{\boldsymbol{j}}\sum_{k=1}^{K}|\textbf{E}(\mathbf{r}_k)+&\textbf{E}_s(\mathbf{r}_k)-s_k|^2\\
		s.t. \sum_{p=1}^{\infty}|j_p|^2&\leq P_T.
	\end{aligned}
\label{equ_q3}
\end{equation}
Since the scattered electric fields are solved using numerical calculation methods, the analytic relationship between $\textbf{E}_s(\textbf{r}_k)$ and $\boldsymbol{j}_{opt}$ cannot be obtained. Therefore, an iterative approach is designed to solve $\mathcal{P}2$, as depicted in Algorithm~\ref{alg:alg1}. Firstly, assume that $\textbf{E}_s(\mathbf{r}_k)=\textbf{0}$, obtain $\boldsymbol{j}_{opt}$ using the proposed iterative method and calculate $\textbf{E}_s(\mathbf{r}_k)$ by using (\ref{equ_sca}). Then, substituting $\textbf{E}_s(\mathbf{r}_k)$ into (\ref{equ_q3}) to calculate $\boldsymbol{j}_{opt}$ and determine whether the convergence condition has been reached. Similarly, the computational complexity of this method depends on (\ref{equ_36}), which is also $O(K^3)$.


\begin{algorithm}[H]
	\renewcommand{\algorithmicrequire}{\textbf{Input:}}
	\renewcommand{\algorithmicensure}{\textbf{Output:}}
	\caption{An iterative method for solving $\boldsymbol{j}_{opt}^0$.}
	\label{alg:alg1}
	\begin{algorithmic}[1]
		\REQUIRE $\varepsilon_1$ when $\textbf{E}_s(\mathbf{r}_k)=\textbf{0}$.
		\STATE Calculate $\boldsymbol{j}_{opt}^0$ by (\ref{equ_36}).
		\STATE \textbf{Initialization:} $i=1$.
		\STATE \textbf{repeat}
		\STATE \hspace{0.5cm}$i=i+1$.
		\FORALL{$|\boldsymbol{j}_{opt}^i-\boldsymbol{j}_{opt}^{i-1}|\textless\varepsilon_1|\boldsymbol{j}_{opt}^{i-1}|$}
		\STATE Calculate $\textbf{E}_s(\mathbf{r}_k)$ by (\ref{equ_sca}).
		\STATE $s_k=s_k-(\textbf{E}(\mathbf{r}_k)+\textbf{E}_s(\mathbf{r}_k))$.
		\STATE Calculate $\boldsymbol{j}_{opt}^i$ by (\ref{equ_36}).
		\ENDFOR
		\STATE \textbf{until} $|\boldsymbol{j}_{opt}^i|=|\boldsymbol{j}_{opt}^{i-1}|$.
		\ENSURE $\boldsymbol{j}_{opt}^i$
	\end{algorithmic}
	\label{alg1}
\end{algorithm}

\subsection{A 3D Continuous-Space Electromagnetic Channel Model}
In 3D continuous space, the electric fields are related to the sampling interval when considering the discretization of the Tx and Rx volumes. After discretization, the electric fields in 3D continuous-space follow\cite{Liwei}
\begin{equation}
	\begin{aligned}
		&\textbf{E}(\textbf{r})=\sum_{n=1}^{N_t}\int_{\delta_t}\overline{\textbf{G}}(\textbf{r},\textbf{r}'_n)\textbf{J}(\textbf{r}'_n)d^3\textbf{r}_n'\\&=\sum_{n=1}^{N_t}\int_{-\frac{\delta_{t}^x}{2}}^{\frac{\delta_{t}^x}{2}}\int_{-\frac{\delta_{t}^y}{2}}^{\frac{\delta_{t}^y}{2}}\int_{-\frac{\delta_{t}^z}{2}}^{\frac{\delta_{t}^z}{2}}\left(\overline{\textbf{I}}+\frac{\nabla\nabla}{k^2}\right)\frac{e^{ikr_n}}{4\pi r_n}\textbf{J}(\textbf{r}'_n)dx'dy'dz'
	\end{aligned}
\end{equation}
where $N_t$ is the number of sample points at the source region, $r_n=|\textbf{r}-\textbf{r}'_n|$ is the distance between the Tx and Rx point, $\delta_t=\delta_{t}^x\delta_{t}^y\delta_{t}^z$ is the sample volume at the source region, and $\delta_{t}^x$, $\delta_{t}^y$, and $\delta_{t}^z$ are sample lengths in the directions of $x$, $y$, and $z$, respectively.

It is difficult to measure the statistical properties of the proposed continuous-space channel model, therefore, they will be simplified using a numerical method. When analyzing complex multipath channels using dyadic Green's functions, approximate numerical solutions to eigenfunctions are employed. The channel impulse response between the $n$th sample point at the Tx and the $m$th sample point at the Rx can be represented as\cite{Liwei}
\begin{equation}
	\begin{aligned}
		\textbf{H}_{mn}(t)=&\delta_t\delta_r\text{sinc}\left(\frac{k(x_m-x_{n}')\delta_t^x}{2r_{mn}(t)}\right)\text{sinc}\left(\frac{k(y_m-y_{n}')\delta_t^y}{2r_{mn}(t)}\right)\\&\text{sinc}\left(\frac{k(z_m-z_{n}')\delta_t^z}{2r_{mn}(t)}\right)\overline{\textbf{G}}_{mn}(t)\textbf{J}(\textbf{r}'_n)
	\end{aligned}
\label{general}
\end{equation}
where $\delta_r$ is the sample volume at the Rx side,  $r_{mn}(t)=|\textbf{r}_m(t)-\textbf{r}_n'(t)|$ is the distance between the $m$th sample point at the Rx and the $n$th sample point at the Tx, $\text{sinc}(x)=\sin(x)/x$ is the sinc function, and $x_m-x_{n}'$, $y_m-y_{n}'$, and $z_m-z_{n}'$ are the differences in distance between the $n$th Tx and the $m$th Rx sample areas on the $x$, $y$, and $z$ axes, respectively. The dyadic Green's function $\overline{\textbf{G}}_{mn}(t)$ between the $n$th Tx and the $m$th Rx sample points is expressed as
\begin{equation}
	\begin{aligned}
		&\overline{\textbf{G}}_{mn}(t)=\left[\left(1+\frac{i}{kr_{mn}(t)}-\frac{1}{k^2r_{mn}^2(t)}\right)\overline{\textbf{I}}\right.
		\\&\left.-\left(1+\frac{3i}{kr_{mn}(t)}-\frac{3}{k^2r_{mn}^2(t)}\right)\hat{\textbf{r}}_{mn}\hat{\textbf{r}}_{mn}\right]\frac{e^{ikr_{mn}(t)}}{4\pi r_{mn}(t)}
	\end{aligned}
\end{equation}
where $\hat{\textbf{r}}_{mn}$ denotes the direction between the $n$th Tx and the $m$th Rx sample area.

For far-field channel models, the dyadic Green's function in (\ref{general}) can be simplified as
\begin{equation}
	\overline{\textbf{G}}_{mn}(t)|_F=\frac{1}{4\pi r_{mn}(t)}\left(\overline{\textbf{I}}-\hat{\textbf{r}}_{mn}\hat{\textbf{r}}_{mn}\right)e^{ikr_{mn}(t)}.
\end{equation}

\section{Statistical Properties and Channel Capacities}
In this section, statistical properties, single-user channel capacities, and multi-user channel capacities of the proposed channel model are derived and calculated.
\subsection{Temporal ACF and Spatial CCF}
The temporal autocorrelation function (ACF) describes the channel correlation coefficient between different time instants. Since the wireless channels are time-variant, the temporal ACF can be calculated as
\begin{equation}
	\textbf{R}_{mn}(\Delta t)=\mathbb{E}\{\textbf{H}_{mn}(t)\textbf{H}^\dagger_{mn}(t+\Delta t)\}
\end{equation}
where $\textbf{H}_{mn}(t+\Delta t)$ is the channel impulse response between the $n$th Tx and the $m$th Rx sample areas at time $t+\Delta t$. 

The spatial cross-correlation function (CCF) is proportional to the electric fields and can be calculated as \cite{Liwei}
\begin{equation}
	\begin{aligned}
		&\textbf{R}_{mn,\tilde{m}\tilde{n}}(\Delta \textbf{r}_t,\Delta \textbf{r}_r)\propto \textbf{E}(\textbf{r}_m)\textbf{E}^\text{T}(\textbf{r}_{\tilde{m}})\\&=\iint \overline{\textbf{G}}(\textbf{r}_m,\textbf{r}_n')\overline{\textbf{G}}^\text{T}(\textbf{r}_{\tilde{m}},\textbf{r}'_{\tilde{n}})\textbf{J}(\textbf{r}'_n)\textbf{J}(\textbf{r}'_{\tilde{n}})d\textbf{r}'_nd\textbf{r}'_{\tilde{n}}
	\end{aligned}
	\label{STF}
\end{equation}
where subscripts $m$ and $\tilde{m}$ are the $m$th and $\tilde{m}$th sample point at the Rx, $n$ and $\tilde{n}$ are the $n$th and $\tilde{n}$th sample point at the Tx, $\Delta \textbf{r}_t$ and $\Delta \textbf{r}_r$ are the distance differences at the Tx and Rx, respectively. The locations of the $m$th and $\tilde{m}$th sample point at the Rx are denoted as $\textbf{r}_m$ and $\textbf{r}_{\tilde{m}}$, respectively. The locations of the $n$th and $\tilde{n}$th sample point at the Tx are denoted as $\textbf{r}'_n$ and $\textbf{r}'_{\tilde{n}}$, respectively.
\subsection{Single-User Channel Capacity}
In single-user communications, assume that each channel is independent, the channel capacity of the continuous-space electromagnetic channel model can be defined as
\begin{equation}
	C=\max\limits_{\boldsymbol{j}}\sum_{p=1}^{\infty}\log_2\left(1+\frac{\sigma_p^2|j_p|^2}{N}\right)
\end{equation} 
where $\sigma_p$ is the channel gain, $j_p$ is the source current, and $N$ is the additive white Gaussian noise (AWGN). In addition, the power constraint needs to be considered in the channel capacity calculation and the optimization problem is built as
\begin{equation}
	\begin{aligned}
		\mathcal{P}3: \max\limits_{\boldsymbol{j}}\sum_{p=1}^{\infty}\log_2&\left(1+\frac{\sigma_p^2|j_p|^2}{N}\right)\\
		s.t. \sum_{p=1}^{\infty}&|j_p|^2\leq P_T
	\end{aligned}
\end{equation}
which can be solved by water-filling algorithm and the solution is \cite{ref_waterfilling}
\begin{equation}
	|(j_p)_{opt}|^2=\max\left[wl-\frac{N}{\sigma_p^2},0\right]
\end{equation}
where $\max[a,b]$ denotes the larger value in $a$ and $b$, and $wl$ is the water level. The optimal solution of $j_p$ is $(j_p)_{opt}$, which follows that
\begin{equation}
	\sum_{p}|(j_p)_{opt}|^2=P_T.
\end{equation}
\subsection{Multi-User Channel Capacity}
The channel capacity of the continuous-space electromagnetic channel model for multi-user communications can reflect the maximum transmission rate in wireless channels. The channel capacity of the proposed channel model, taking scatterers into account, can be calculated as
\begin{equation}
	C=\sum_{k=1}^{K}\log_{2}\left(1+\frac{|\textbf{E}(\textbf{r}_k)+\textbf{E}_s(\textbf{r}_k)|^2}{N}\right)
\end{equation}
where $K$ is the total number of users. To obtain the relationship between the channel capacity and transmit power, we can use the Monte-Carlo simulation to get its mean value. Therefore, the channel capacity can be calculated as
\begin{equation}
	C=\underset{\boldsymbol{s},\textbf{r}_k}{\mathbb{E}}\left\{\sum_{k=1}^{K}\log_{2}\left(1+\frac{|\textbf{E}(\textbf{r}_k)+\textbf{E}_s(\textbf{r}_k)|^2}{N}\right)\right\}
\end{equation}
where $\textbf{r}_k$ is the user location and $\boldsymbol{s}$ is the transmit signal vector. The $k$th transmit signal is $s_k=e^{i\phi_k}$, where $\phi_k$ follows a uniform distribution within the range of $[0,2\pi)$.

When the interference is considered in multi-user communications, the channel capacities are determined by employing  precoding schemes, which can be calculated as
{\small\begin{equation}
	\begin{aligned}
		&C=\underset{\boldsymbol{s},\textbf{r}_k}{\mathbb{E}}\left\{\sum_{k=1}^{K}\log_{2}\left(1+\frac{}{}\right.\right.\\
		&\left.\left.\frac{\left(\textbf{E}(\textbf{r}_k)+\textbf{E}_s(\textbf{r}_k)\right)\mathbf{W}_k\mathbf{W}^\dagger_k(\textbf{E}(\textbf{r}_k)+\textbf{E}_s(\textbf{r}_k))^\dagger}{\sum_{u=1,u\neq k}^{K}\left(\textbf{E}(\textbf{r}_k)+\textbf{E}_s(\textbf{r}_k)\right)\mathbf{W}_u\mathbf{W}^\dagger_u(\textbf{E}(\textbf{r}_k)+\textbf{E}_s(\textbf{r}_k))^\dagger+N\mathbf{I}}\right)\right\}
	\end{aligned}
\end{equation}}
where $\mathbf{W}_k$ is the precoding matrix of the $k$th user.

\section{Results and Analysis}
In this section, channel statistical properties, single-user channel capacities, and multi-user channel capacities of the continuous-space electromagnetic channel model are simulated and verified by the full-wave simulation software. 
\begin{figure}[tb]
	\centering
	\includegraphics[width=0.34\textheight]{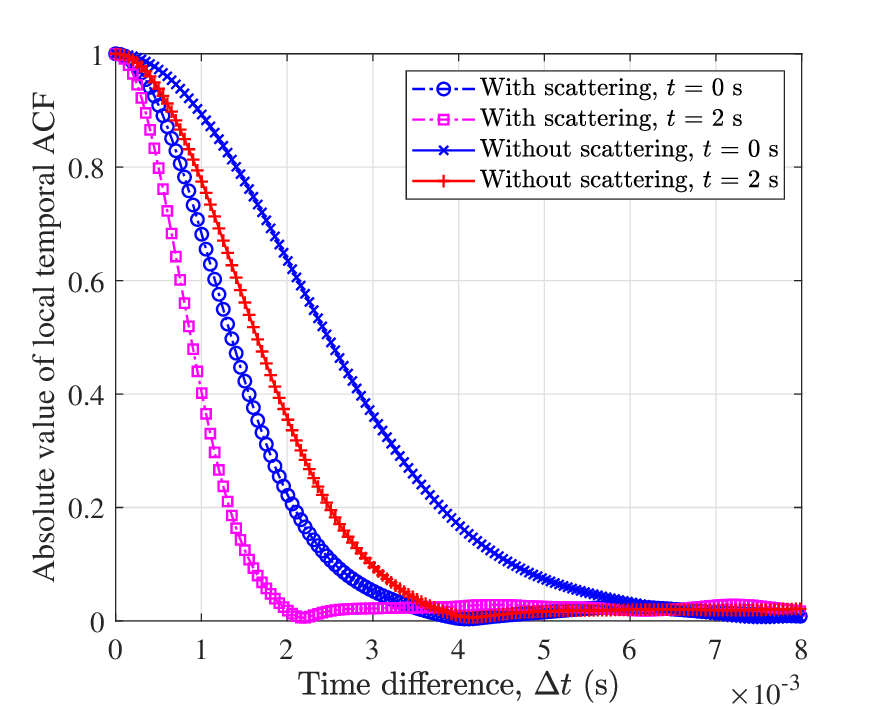}
	\caption{Temporal autocorrelation functions of the continuous-space electromagnetic channel model with and without scattering.}
	\label{Fig16}
\end{figure}
\begin{figure}[tb]
	\centering
	\includegraphics[width=0.34\textheight]{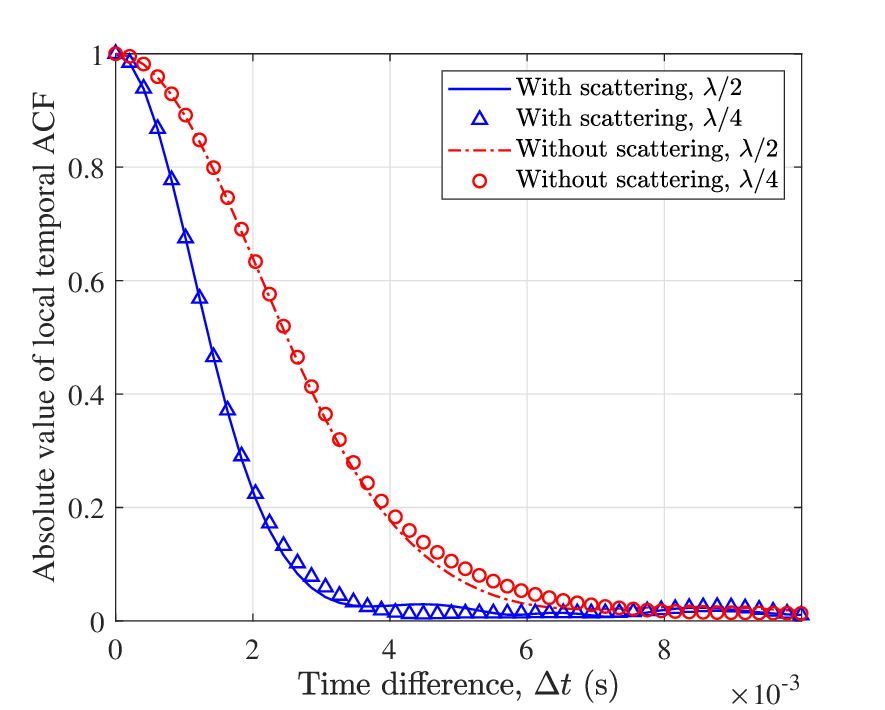}
	\caption{Temporal autocorrelation functions of the continuous-space electromagnetic channel model for different sample intervals.}
	\label{Fig17}
\end{figure}
\subsection{Statistical Properties}
Assume that the carrier frequency $f_c$ is 30 GHz, the wavelength $\lambda$ is 0.01 m. Since the MoM utilized in the proposed channel model results in a high computational complexity, the radius of the transmit sphere is set to $2\lambda$ and the radius of the received sphere is set to $20\lambda$ in the simulation. However, the apertures of the Tx and Rx antennas can be adjusted to the sizes envisioned for extremely large MIMO schemes. In addition, the range of transmit power $P_T$ is 30--50 dBm. In Fig.~\ref{Fig16}, simulation result shows that the temporal ACF at $t$ = 0 s and $t$ = 2 s are different, which demonstrates the temporal non-stationarity of the proposed channel model. The temporal ACF of the proposed channel with scattering is lower than that without scatterers, that is, scatterers can accelerate the decrease of channel correlation at different time instants. This is because scatterers can enhance the multipath effect in wireless channel models. In Fig.~\ref{Fig17}, temporal ACFs for different sample intervals are shown. It is observed that the temporal ACF with $\lambda/2$ sample interval is almost the same as the temporal ACF with $\lambda/4$ sample interval. This is because when the sample interval reaches $\lambda/2$, the temporal ACF can already reflect the full information of channels according to the Nyquist sampling theorem. The temporal ACF will not be affected when further decreasing the sample interval.

\begin{figure}[tb]
	\centering
	\includegraphics[width=0.34\textheight]{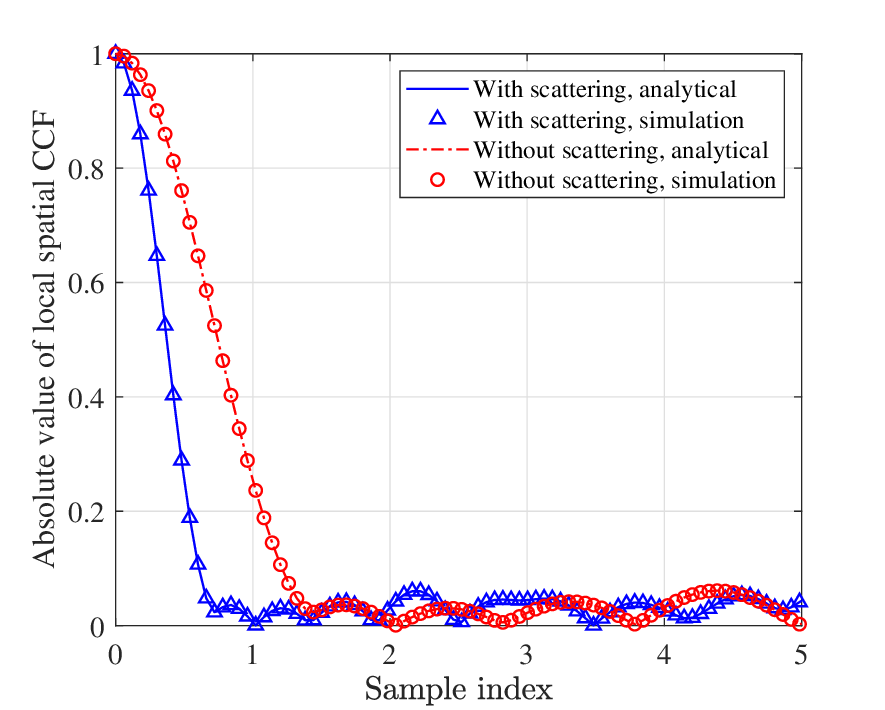}
	\caption{The spatial cross-correlation functions of the continuous-space electromagnetic channel model with and without scattering.}
	\label{Fig14}
\end{figure}
\begin{figure}[tb]
	\centering
	\includegraphics[width=0.34\textheight]{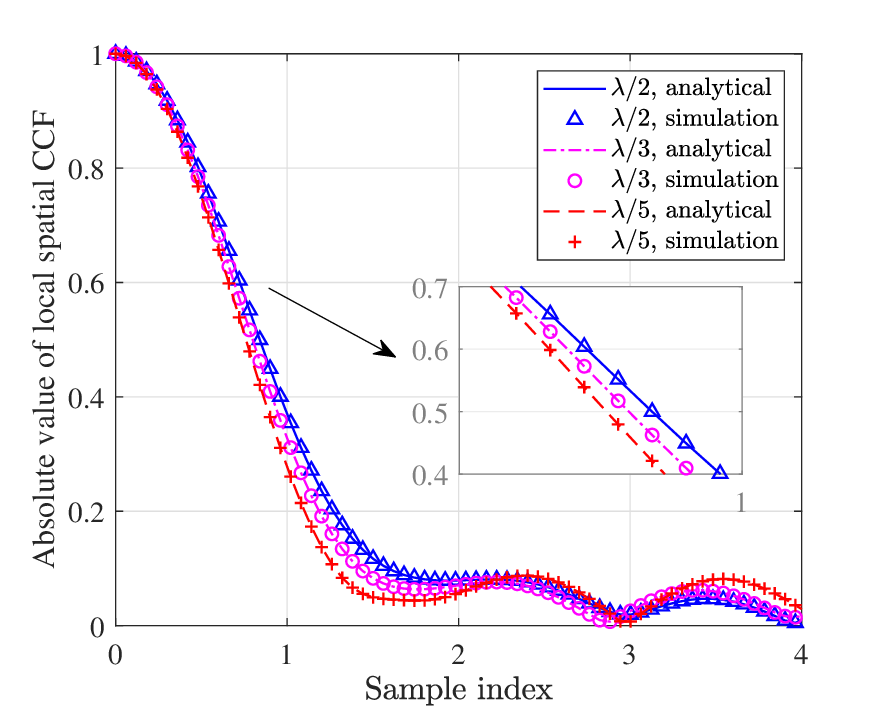}
	\caption{The spatial cross-correlation functions of the continuous-space electromagnetic channel model with different sample intervals.}
	\label{Fig15}
\end{figure}

\begin{figure*}[tb]
	\centering
	\subfloat[$|\mathbf{E}_{\mathbf{r}}|$ in the $\theta$ direction.]{
		\includegraphics[width=0.2\textheight]{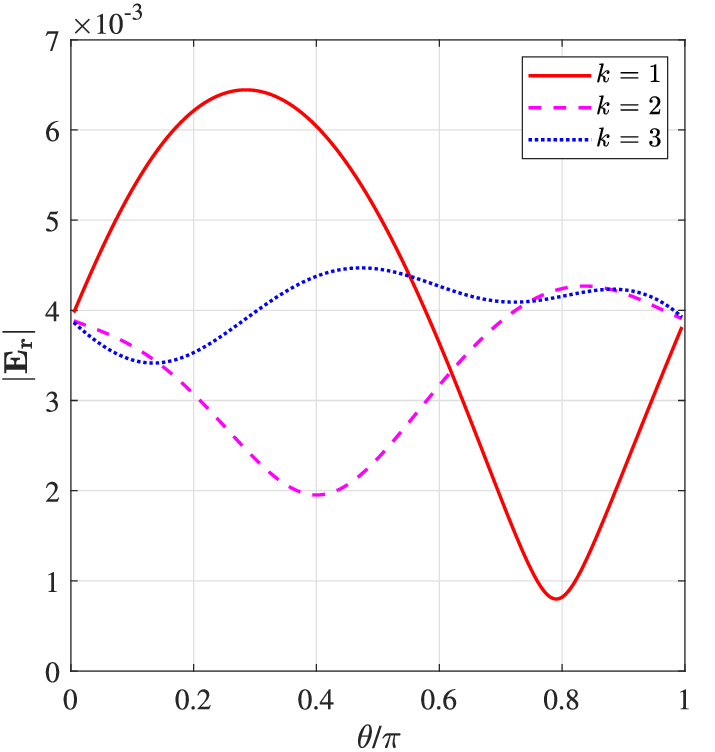}}
	\hfil
	\subfloat[$|\mathbf{E}_{\theta}|$ in the $\theta$ direction.]{
		\includegraphics[width=0.2\textheight]{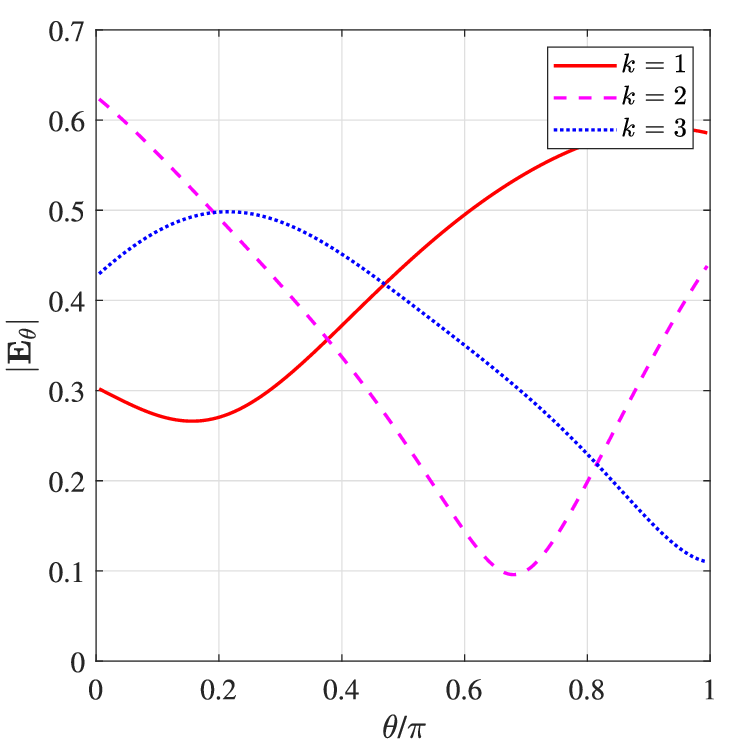}}
	\hfil
	\subfloat[$|\mathbf{E}_{\varphi}|$ in the $\theta$ direction.]{
		\includegraphics[width=0.2\textheight]{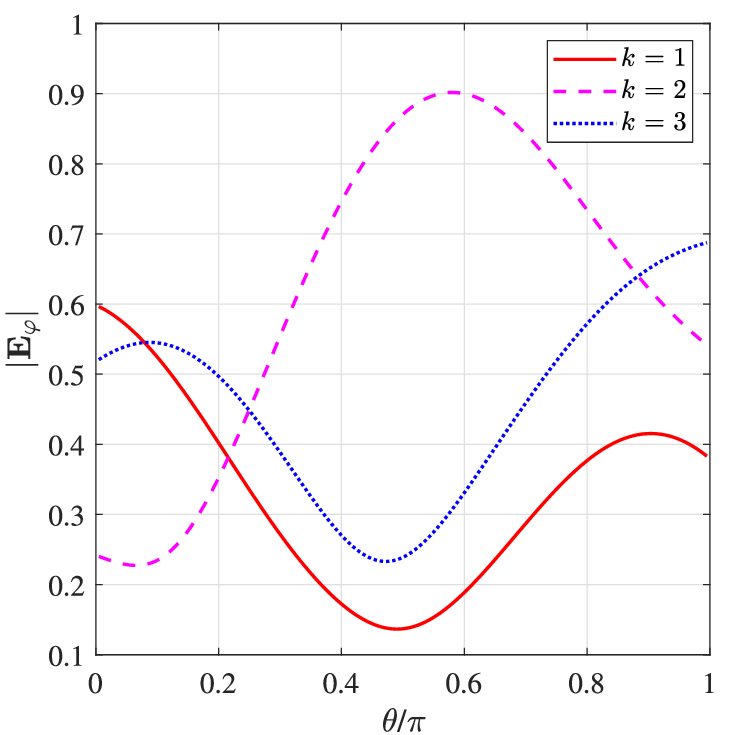}}
	\hfil
	\subfloat[$|\mathbf{E}_{\mathbf{r}}|$ in the $\phi$ direction.]{
		\includegraphics[width=0.2\textheight]{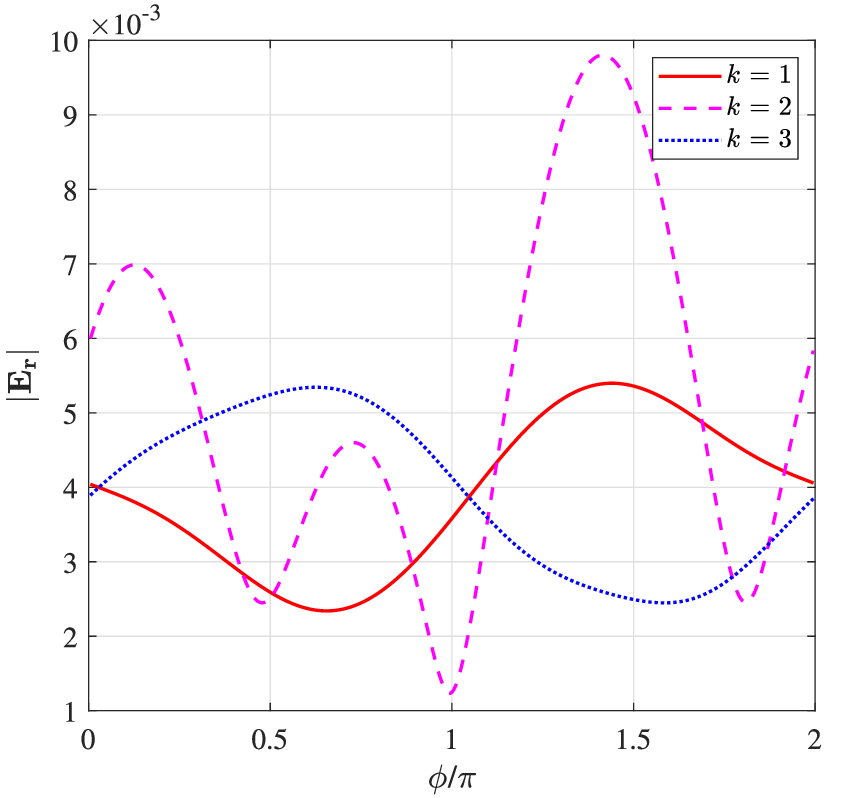}}
	\hfil
		\subfloat[$|\mathbf{E}_{\theta}|$ in the $\phi$ direction.]{
		\includegraphics[width=0.2\textheight]{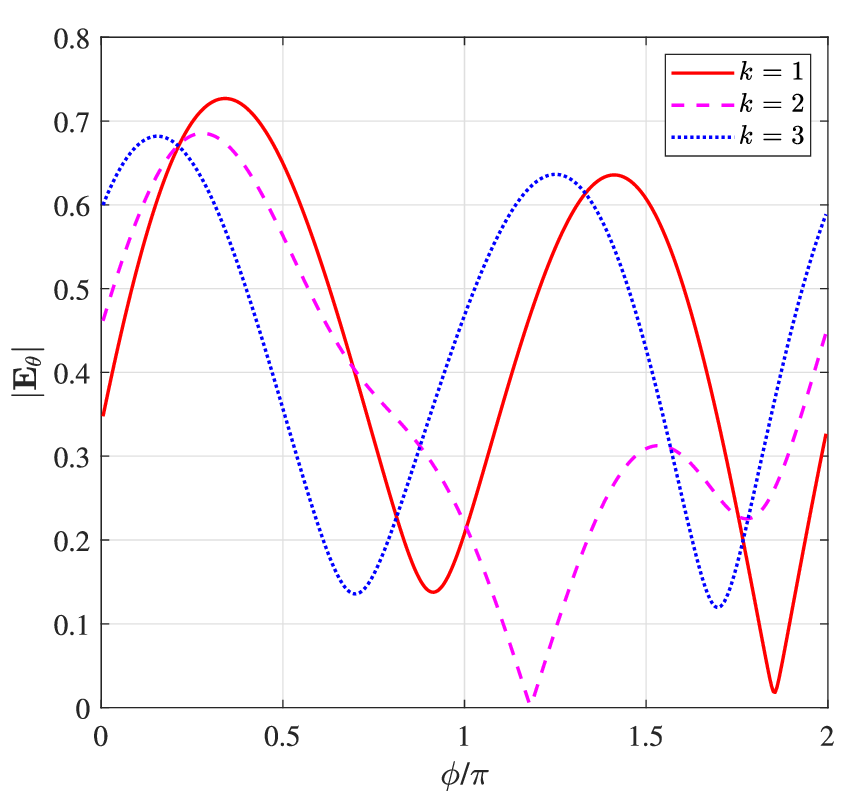}}
	\hfil
		\subfloat[$|\mathbf{E}_{\varphi}|$ in the $\phi$ direction.]{
		\includegraphics[width=0.2\textheight]{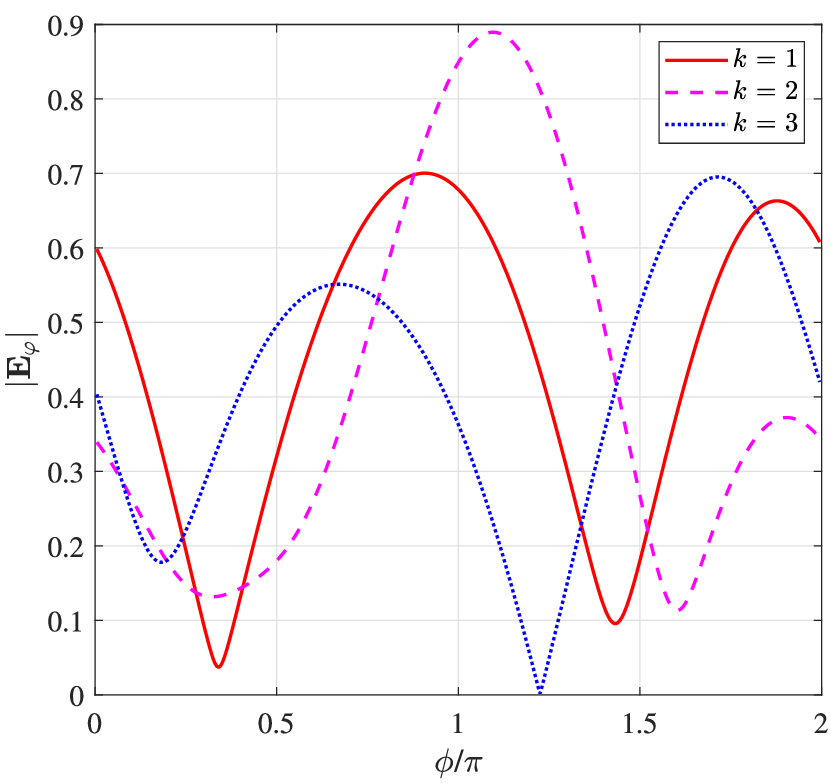}}
	\caption{Radiation patterns corresponding to the source current $\textbf{J}(\textbf{r})$ of the multi-user communication in (a)-(c) the $\theta$ direction and (d)-(f) the $\phi$ direction.}
	\label{Fig18}
\end{figure*}
\begin{figure}[tb]
	\centering
	\includegraphics[width=0.33\textheight]{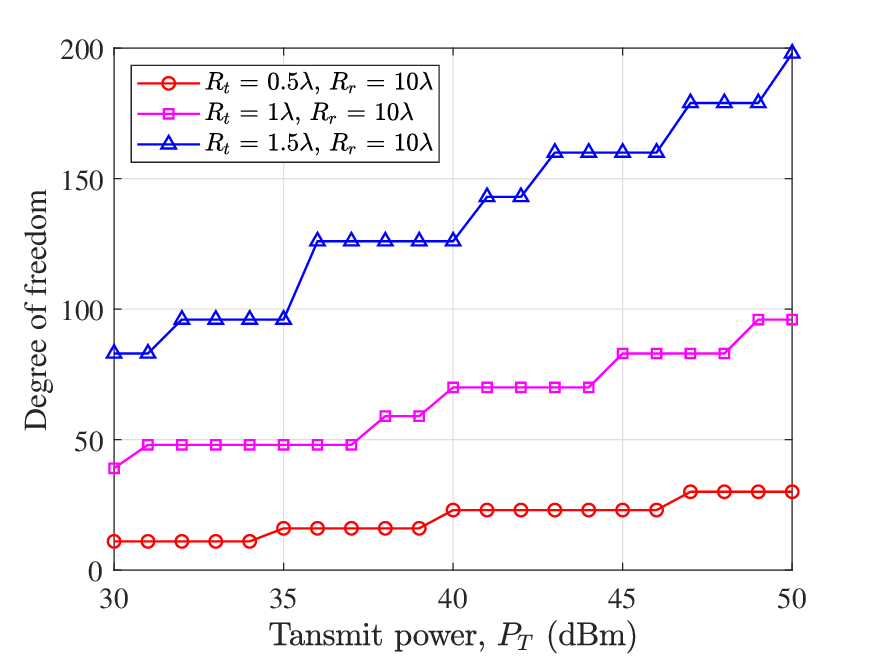}
	\caption{The degree-of-freedom of the continuous-space electromagnetic channel model for different apertures of the Tx ($R_t=0.5 \lambda, 1 \lambda, 1.5 \lambda, R_r=10 \lambda, f=30\ \text{GHz}, \text{and}\ D=10\ \text{m}$).}
	\label{Fig4}
\end{figure}
\begin{figure}[tb!]
	\centering
	\includegraphics[width=0.33\textheight]{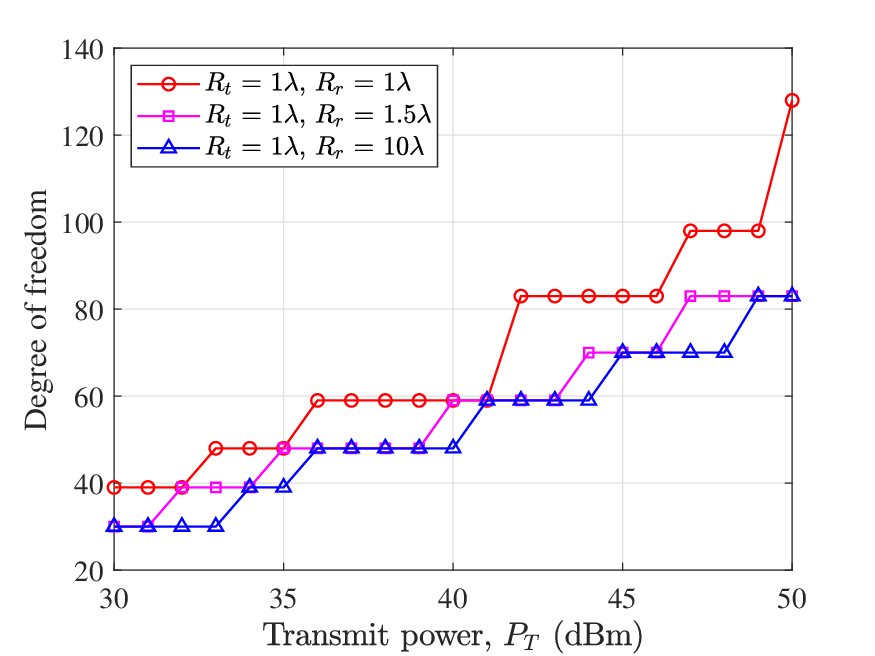}
	\caption{The degree-of-freedom of the continuous-space electromagnetic channel model for different apertures of the Rx ($R_t=0.5 \lambda, R_r=1\lambda,1.5\lambda, 10 \lambda, f= 30\ \text{GHz}, \text{and}\ D=10\ \text{m}$).}
	\label{Fig5}
\end{figure}
\begin{figure}[tb]
	\centering
	\includegraphics[width=0.33\textheight]{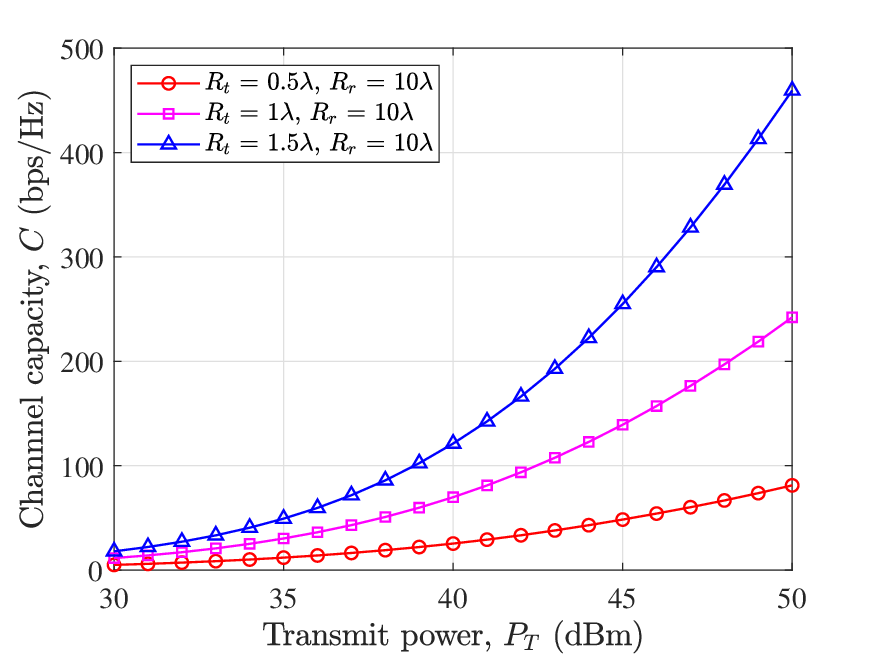}
	\caption{Channel capacity of the continuous-space electromagnetic channel model for different apertures of the Tx ($R_t=0.5 \lambda, 1 \lambda, 1.5 \lambda, R_r=10 \lambda, f=30\ \text{GHz}, \text{and}\ D=10\ \text{m}$).}
	\label{Fig6}
\end{figure}

\begin{figure}[tb]
	\centering
	\includegraphics[width=0.33\textheight]{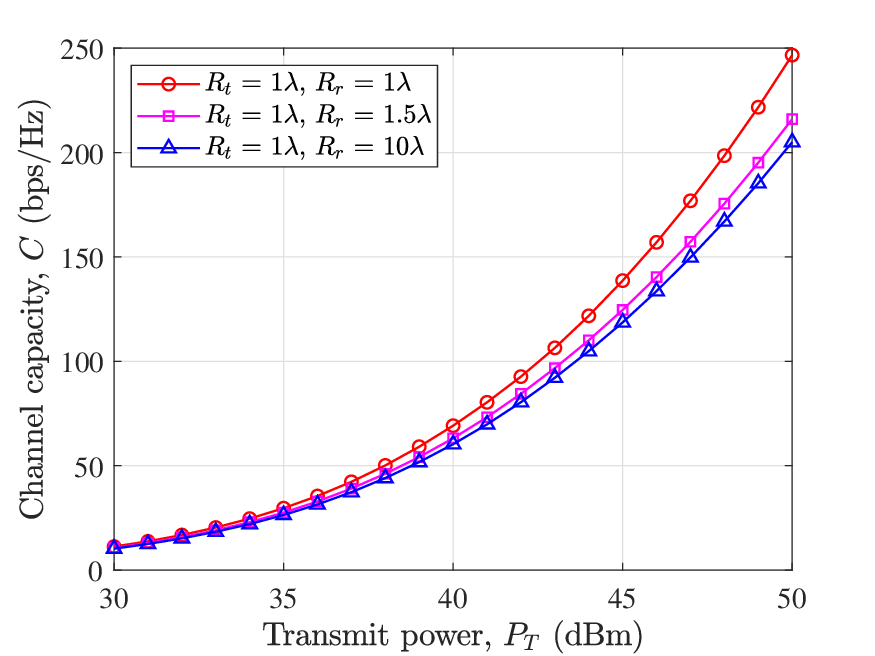}
	\caption{Channel capacity of the continuous-space electromagnetic channel model for different apertures of the Rx ($R_t=0.5 \lambda, R_r=1\lambda,1.5\lambda, 10 \lambda, f= 30\ \text{GHz}, \text{and}\ D=10\ \text{m}$).}
	\label{Fig7}
\end{figure}

Spatial CCFs of the proposed channel model with and without scattering are shown in Fig.~\ref{Fig14}. When scattering is considered, the spatial CCF decreases rapidly. This is due to the multipath effect in a rich-scattering environment, which can reduce the correlation between different positions. In addition, the analytical results align with the simulation results, confirming the validity of the theoretical derivation and the accuracy of the proposed channel model. The spatial CCFs with different sample intervals are shown in Fig.~\ref{Fig15}. The sample interval in the continuous-space channel can be considered as the antenna spacing in the discrete-space channel. Since the antenna spacing is usually around half a wavelength in 5G communication systems, while holographic MIMO communications are new technologies in 6G communication systems, sample intervals less than half a wavelength are chosen to simulate the spatial CCFs. Simulation results show that spatial CCFs decrease as the sample interval decreases. However, the impact of varying sample intervals on spatial CCFs is not readily apparent, primarily due to the small volumes of the Tx and Rx spheres. 
\subsection{Antenna Patterns}
After solving the optimization problem $\mathcal{P}$2, the source current coefficient $\boldsymbol{j}$ that can enable multi-user communications can be obtained. The radiation pattern corresponding to the source current $\textbf{J}(\textbf{r})$ can be calculated using (\ref{equ_SVD}). In Fig.~\ref{Fig18}, the radiation patterns corresponding to the source current $\textbf{J}(\textbf{r})$ of multi-user communications are provided, which can offer guidance for antenna synthesis in practice\cite{CSEM-2008}. Assume that there are three users in the communication network, where the radius of the source region is $\lambda$ and the radius of the receiving region is $20\lambda$. The radiation patterns in the $\theta$ and $\phi$ directions are shown in Figs.~\ref{Fig18}(a)-(c) and Figs.~\ref{Fig18}(d)-(f), respectively, where the components of the electric fields in the $r$, $\theta$, and $\varphi$ directions are represented. In practice, this paper provides two methods to achieve the same performance in multi-user communications. The first one is to design the source current $\textbf{J}(\textbf{r})$ calculated based on optimization problem $\mathcal{P}$2, while the second one is to conduct antenna synthesis according to the radiation pattern given in Fig.~\ref{Fig18}. Therefore, the continuous-space electromagnetic channel model can guide the antenna synthesis and communication system design.

\subsection{Single-User Communication Scenario}
In the single-user communication scenario, we studied the DoFs and channel capacities of the continuous-space electromagnetic channel model in free space.

In Fig.~\ref{Fig4}, it is evident that the DoF increases as the transmit power increases when the apertures of the Tx and Rx are fixed. When the transmit power is fixed, the DoF increases as the source region expands. In Fig.~\ref{Fig5}, it also shows that DoF increases as the transmit power increases. When the transmit power is fixed, the DoF increases as the receiving region decreases. This is because when the source region increases or the receiving region decreases, the channel gains $\sigma_p$ will increase. The influence of the source region on DoF is much larger than that of the receiving region. Especially, the influence of the receiving region on DoF is more pronounced obvious when the receiving region is smaller.

In Figs.~\ref{Fig6} and \ref{Fig7}, the channel capacities of the continuous-space electromagnetic channel model for different apertures of the Tx and Rx are shown. In Fig.~\ref{Fig6}, the experimental results show that when the transmit power is fixed, the channel capacity increases as the aperture of the source region increases. When the apertures of the Tx and Rx are fixed, the channel capacity will increase as the transmit power increases. It is consistent with the trend of the spatial DoF. In Fig.~\ref{Fig7}, the channel capacity of the continuous-space electromagnetic channel model shows little difference for different apertures of the Rx. When the aperture of the Rx increases, the channel capacity will decrease. This is because the channel capacity and the spatial DoF are determined by the channel gain $\sigma_p$ and the transmit power $P_T$.
\subsection{Multi-User Communication Scenario}
In a multi-user communication scenario, the item number $P$ of the SVD needs to be chosen properly. This can be determined by the minimum mean squared error (MMSE) in (\ref{err}) and the transmit power $P_T$. Firstly, we studied the relationship between the SVD items $P$ and the signal error $err$ for different numbers of users. 
\begin{figure}[tb]
	\centering
	\includegraphics[width=0.34\textheight]{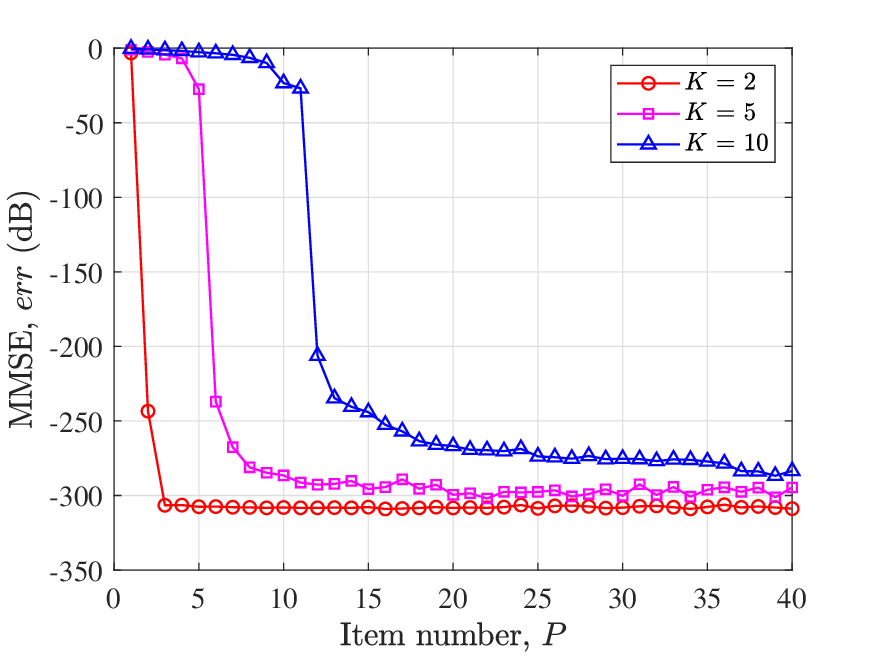}
	\caption{The relationship between minimum mean squared error and the item number of SVD for different numbers of users.}
	\label{Fig8}
\end{figure}
\begin{figure}[tb]
	\centering
	\includegraphics[width=0.34\textheight]{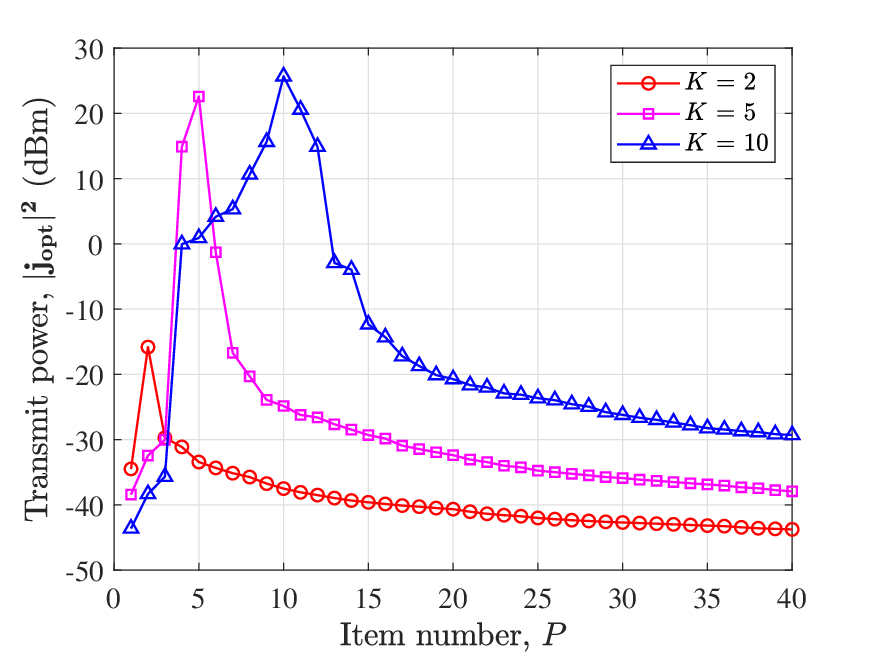}
	\caption{The relationship between the transmit power and the item number of SVD for different numbers of users.}
	\label{Fig9}
\end{figure}

In Fig.~\ref{Fig8}, the MMSE for different numbers of users is simulated. It is shown that when the item numbers of SVD $P<K$, the MMSE will decrease as the value of $P$ increases. When the item numbers of SVD $P=K$, the MMSE decreases rapidly as $P$ increases and stabilizes after reaching a very small value. Therefore, it is appropriate to choose $P>K$. In Fig.~\ref{Fig9}, for different numbers of users, the transmit power $|\boldsymbol{j}_{opt}|^2$ initially increases, then decreases with the increase of the item value $P$, and finally tends to be stable. It is shown that when $P=K$, the transmit power reaches its maximum value. When $P$ equals $3K$, the transmit power decreases to a relatively low level and remains unchanged. In real communication scenarios, the less energy consumed by the source region, the more energy-efficient communication becomes. Therefore, $P=3K$ is the most appropriate choice when jointly considering the influences of $P$ on the MMSE and transmit power, as well as the complexity of the multi-user communications setup.
\begin{figure}[tb]
	\centering
	\includegraphics[width=0.33\textheight]{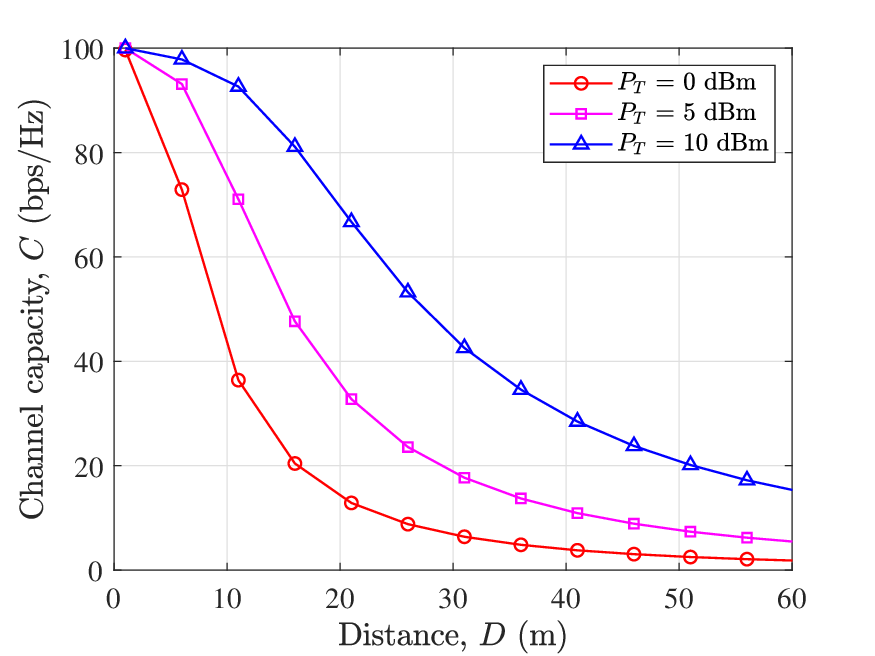}
	\caption{The channel capacity of the continuous-space electromagnetic channel model for different distance $D$.}
	\label{Fig10}
\end{figure}
\begin{figure}[tb]
	\centering
	\includegraphics[width=0.33\textheight]{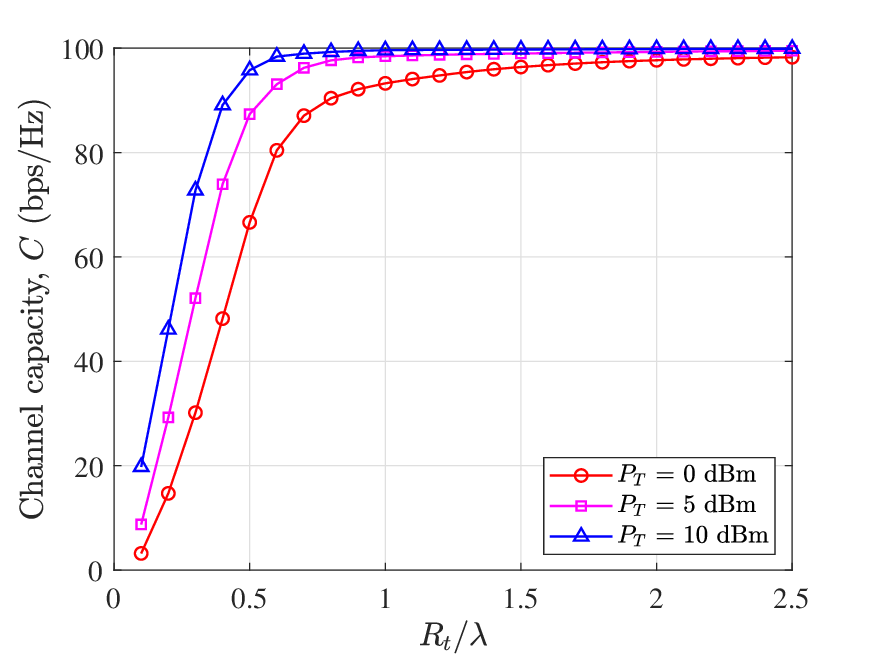}
	\caption{The channel capacity of the continuous-space electromagnetic channel model for different apertures of the Tx.}
	\label{Fig11}
\end{figure}

\begin{figure}[tb]
	\centering
	\includegraphics[width=0.33\textheight]{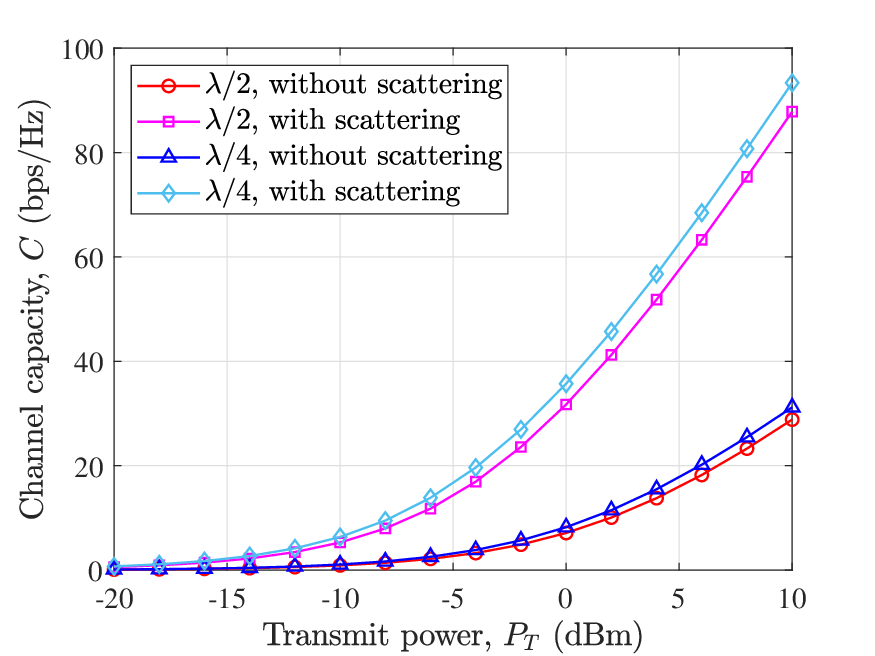}
	\caption{Channel capacities of the continuous-space electromagnetic channel model versus transmit power with and without scattering for different sample intervals.}
	\label{Fig12}
\end{figure}

\begin{figure}[tb]
	\centering
	\includegraphics[width=0.33\textheight]{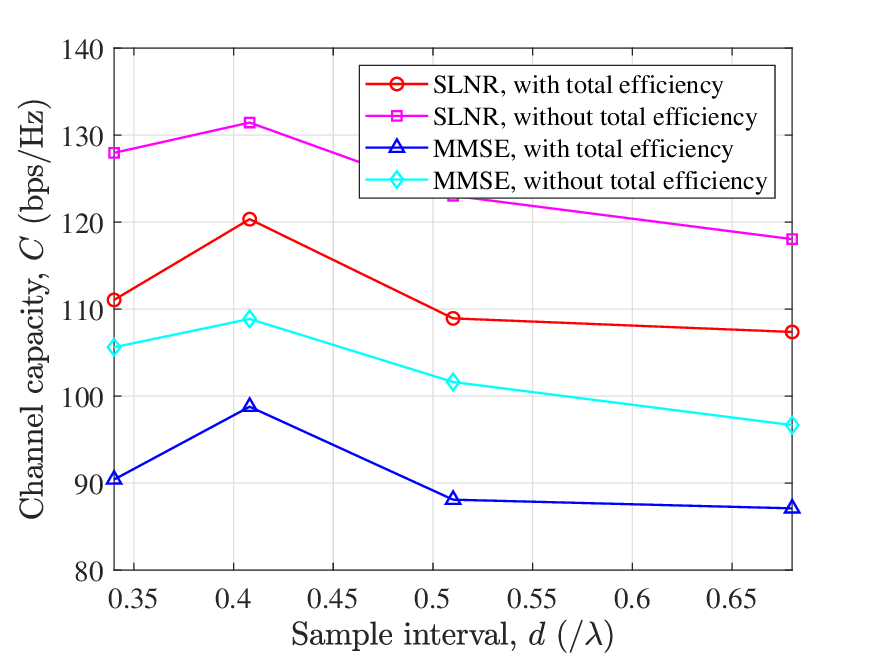}
	\caption{Multi-user channel capacities of continuous-space electromagnetic channel model with different sample intervals.}
	\label{Fig13}
\end{figure}

In the following simulations, there are 10 users and the dimension of the SVD is 30. Assume that the radius of the transmitting sphere is $2\lambda$ and the radius of the receiving sphere is $20\lambda$. In Fig.~\ref{Fig10}, the channel capacity decreases as the distance between the Tx and Rx increases, which is due to the fact that smaller distances result in less attenuation of the electric fields. In Fig.~\ref{Fig11}, the channel capacity increases as the radius of the transmit sphere increases. In addition, it is obvious that the channel capacity increases as the transmit power increases. This is because a larger antenna aperture corresponds to stronger spatial filtering ability, resulting in a higher channel capacity. In Fig.~\ref{Fig12}, multi-user channel capacities of the proposed channel model with and without scattering are compared for different sample intervals. It can be seen that the channel capacity can be increased by about 3~bps/Hz at $P_T$ = 5 dBm when the sample interval is reduced from half a wavelength to $\lambda/4$, and the channel capacity with scattering is approximately three times greatly than that without scattering. It is illustrated that the channel capacity considering scatterers is greater than that without considering scatterers due to the multipath effect. In addition, the channel capacity can increase to a certain extent when the sample interval decreases.

To better illustrate the numerical results, the influence of the sample interval, the precoding scheme, and the total efficiency on channel capacities are investigated through the full-wave CST STUDIO SUITE$^\circledR$. Here, the sample interval ranges from 0.3$\lambda$ to 0.7$\lambda$ and the transmit power is set to 10 dBm. Two precoding schemes including MMSE and signal-to-leakage and noise ratio (SLNR) are adopted to calculate the multi-user capacities. In CST microwave studio, the antenna pattern, bandwidth, and efficiency are considered. The simulation results are shown in Fig. \ref{Fig13}. It can be observed that as the sample interval decreases, the multi-user channel capacity first increases and then decreases. This is because when the sample interval becomes smaller, the difference in the current distribution of the antenna surface is greater, and the regulation of the current distribution will be more precise. However, the total efficiency will decrease rapidly as the antenna spacing decreases. It can also be observed that the multi-user capacity using SLNR precoding is higher than that using MMSE precoding, indicating that SLNR precoding has better performance. In addition, when the sample interval is small, capacities considering the total efficiency are smaller than those without considering the total efficiency.

\section{Conclusions}
In this paper, we have proposed a 3D continuous-space electromagnetic channel model considering scatterers, non-stationarity, and near-field spherical wavefronts for tri-polarized multi-user communications. The MoM has been utilized to calculate the scattered electric fields, while spherical wave functions have been employed to evaluate the near-field effect. In addition, channel statistical properties, single-user channel capacities, and multi-user channel capacities have been simulated and analyzed. Simulation results have shown that the temporal ACF and spatial CCF decrease when considering scatterers in the proposed channel model. Additionally, multi-user channel capacities increase as the distance between the Tx and Rx decreases. Results have also illustrated that channel capacities increase as the transmitting volume and power increase. The channel capacity with scatterers is about three times greater than that without scatterers. In addition, the sample interval also impacts multi-user communications according to full-wave simulation. The results illustrate that when the sample interval decreases, the channel capacity will first increase and then decrease, which is because of the increase in spatial DoF, differences in surface current distributions, and the rapid decrease in total efficiency. Therefore, the analysis of the 3D continuous-space electromagnetic channel model is significant for network planning in multi-user wireless communications. 

\bibliographystyle{IEEEtran}

\begin{IEEEbiography}[{\includegraphics[width=1in,height=1.25in,clip,keepaspectratio]{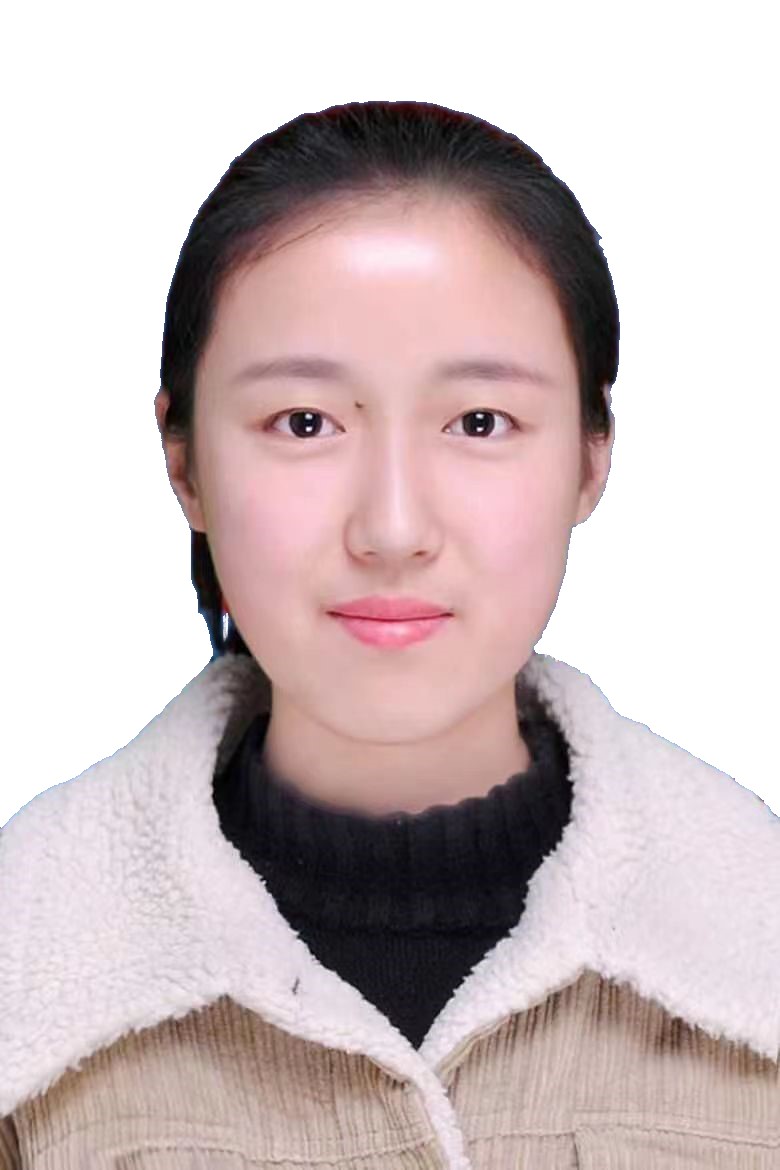}}]{Yue Yang}
	received the B.E. degree in Communication Engineering from Xidian University, China, in 2020. She is currently pursuing the Ph.D. degree in the Nation Mobile
	Communications Research Laboratory, Southeast University, China. Her research interests include electromagnetic information theory, 6G wireless channel modeling and characteristic analysis, and 6G wireless communications.
\end{IEEEbiography}

\begin{IEEEbiography}[{\includegraphics[width=1in,height=1.25in,clip,keepaspectratio]{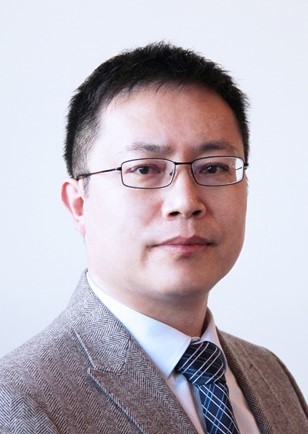}}]{Cheng-Xiang Wang}
	(Fellow, IEEE) received the B.Sc. and M.Eng. degrees in communication and information systems from Shandong University, China, in 1997 and 2000, respectively, and the Ph.D. degree in wireless communications from Aalborg University, Denmark, in 2004. 
	
	He was a Research Assistant with the Hamburg University of Technology, Hamburg, Germany, from 2000 to 2001, a Visiting Researcher with Siemens AG Mobile Phones, Munich, Germany, in 2004, and a Research Fellow with the University of Agder, Grimstad, Norway, from 2001 to 2005. He was with Heriot-Watt University, Edinburgh, U.K., from 2005 to 2018, where he was promoted to a professor in 2011. He has been with Southeast University, Nanjing, China, as a professor since 2018, and he is now the Executive Dean of the School of Information Science and Engineering. He is also a professor with Purple Mountain Laboratories, Nanjing, China. He has authored 4 books, 3 book chapters, and over 600 papers in refereed journals and conference proceedings, including 28 highly cited papers. He has also delivered 31 invited keynote speeches/talks and 18 tutorials in international conferences. His current research interests include wireless channel measurements and modeling, 6G wireless communication networks, and electromagnetic information theory. 
	
	Dr. Wang is a Member of the Academia Europaea (The Academy of Europe), a Member of the European Academy of Sciences and Arts (EASA), a Fellow of the Royal Society of Edinburgh (FRSE), IEEE, IET, and China Institute of Communications (CIC), an IEEE Communications Society Distinguished Lecturer in 2019 and 2020, a Highly-Cited Researcher recognized by Clarivate Analytics in 2017--2020. He is currently an Executive Editorial Committee Member of the IEEE TRANSACTIONS ON WIRELESS COMMUNICATIONS. He has served as an Editor for over ten international journals, including the IEEE TRANSACTIONS ON WIRELESS COMMUNICATIONS, from 2007 to 2009, the IEEE TRANSACTIONS ON VEHICULAR TECHNOLOGY, from 2011 to 2017, and the IEEE TRANSACTIONS ON COMMUNICATIONS, from 2015 to 2017. He was a Guest Editor of the IEEE JOURNAL ON SELECTED AREAS IN COMMUNICATIONS, Special Issue on Vehicular Communications and Networks (Lead Guest Editor), Special Issue on Spectrum and Energy Efficient Design of Wireless Communication Networks, and Special Issue on Airborne Communication Networks. He was also a Guest Editor for the IEEE TRANSACTIONS ON BIG DATA, Special Issue on Wireless Big Data, and is a Guest Editor for the IEEE TRANSACTIONS ON COGNITIVE COMMUNICATIONS AND NETWORKING, Special Issue on Intelligent Resource Management for 5G and Beyond. He has served as a TPC Member, a TPC Chair, and a General Chair for more than 30 international conferences. He received 17 Best Paper Awards from IEEE GLOBECOM 2010, IEEE ICCT 2011, ITST 2012, IEEE VTC 2013 Spring, IWCMC 2015, IWCMC 2016, IEEE/CIC ICCC 2016, WPMC 2016, WOCC 2019, IWCMC 2020, WCSP 2020, CSPS 2021, WCSP 2021, IEEE/CIC ICCC 2022, IEEE ICCT 2023 and IEEE/CIC ICCC 2024.
\end{IEEEbiography}

\begin{IEEEbiography}[{\includegraphics[width=1in,height=1.25in,clip,keepaspectratio]{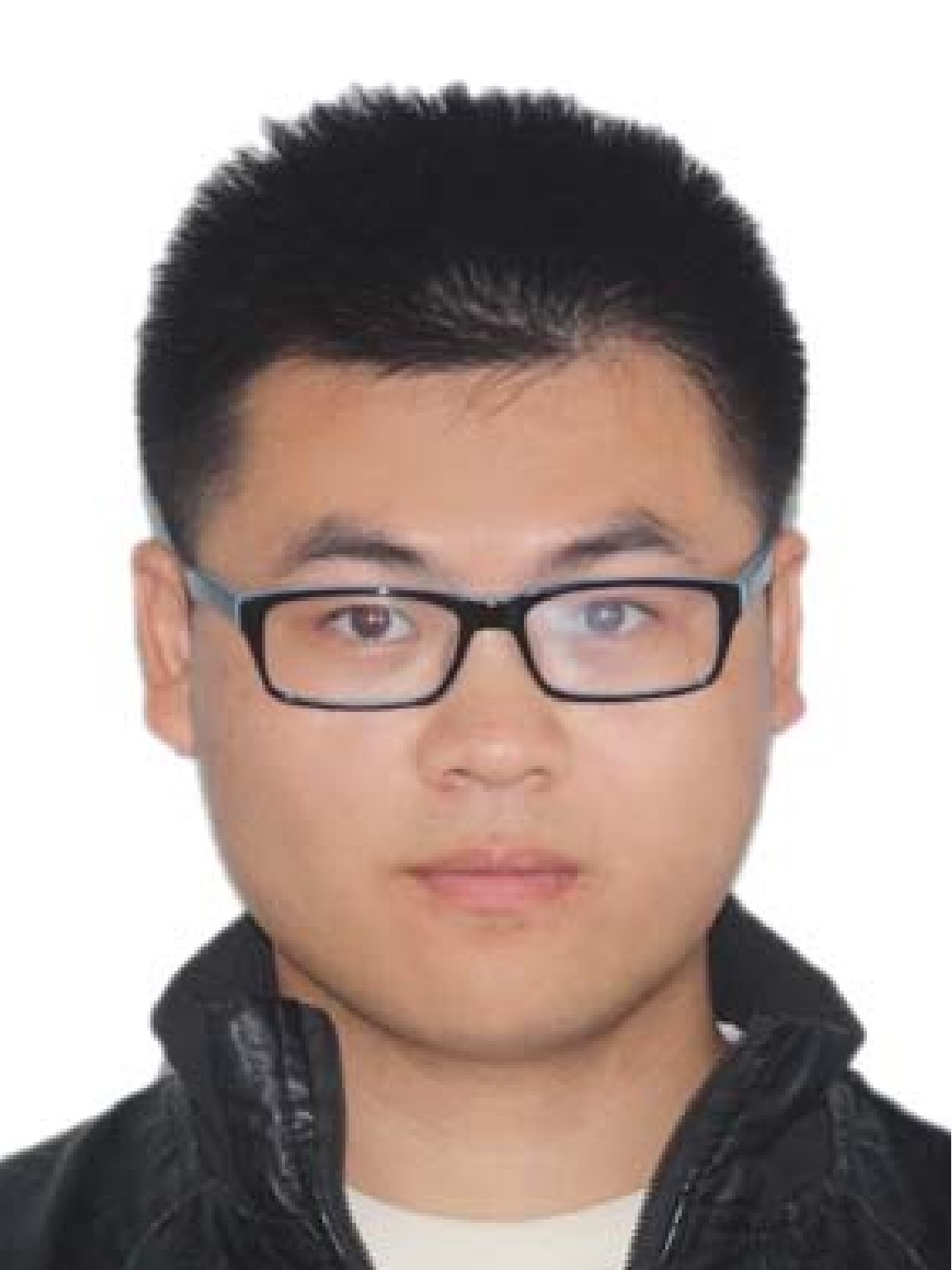}}]{Jie Huang}
	(Member, IEEE) received the B.E. degree in Information Engineering from Xidian University, China, in 2013, and the Ph.D. degree in Information and Communication Engineering from Shandong University, China, in 2018. 
	
	From Oct. 2018 to Oct. 2020, he was a Postdoctoral Research Associate in the National Mobile Communications Research Laboratory, Southeast University, China, supported by the National Postdoctoral Program for Innovative Talents. From Jan. 2019 to Feb. 2020, he was a Postdoctoral Research Associate in Durham University, UK. Since Mar. 2019, he is a part-time researcher in Purple Mountain Laboratories, China. Since Nov. 2020, he is an Associate Professor in the National Mobile Communications Research Laboratory, Southeast University. He has authored and co-authored over 100 papers in refereed journals and conference proceedings. He received the Best Paper Awards from WPMC 2016, WCSP 2020, and WCSP 2021. He has delivered 13 tutorials in IEEE/CIC ICCC 2021, IEEE PIMRC 2021, IEEE ICC 2022, IEEE VTC-Spring 2022, IEEE/CIC ICCC 2022, IEEE VTC-Fall 2022, IEEE PIMRC 2022, IEEE Globecom 2022, IEEE WCNC 2023, IEEE ICC 2023, IEEE/CIC ICCC 2023, IEEE Globecom 2023, and IEEE WCNC 2024. His research interests include millimeter wave, massive MIMO, reconfigurable intelligent surface channel measurements and modeling, wireless big data, electromagnetic information theory, and 6G wireless communications.
\end{IEEEbiography}

\begin{IEEEbiography}[{\includegraphics[width=1in,height=1.25in,clip,keepaspectratio]{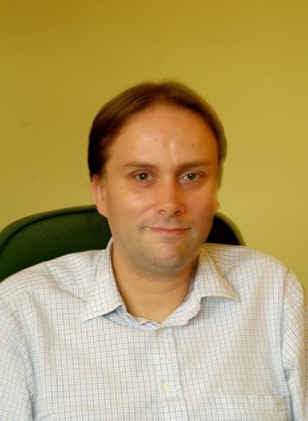}}]{John Thompson}
	(Fellow, IEEE) received the Ph.D. degree in electrical engineering from University of Edinburgh, Edinburgh, U.K., in 1995. 
	
	He currently holds a personal chair in Signal Processing and Communications at the School of Engineering, University of Edinburgh. He currently specializes in antenna array processing, energy-efficient wireless communications and more recently in the application of machine learning to wireless communications. To date, he has published in excess of 500 journal and conference papers on these topics. He is also an Area Editor for the wireless communications topic in IEEE Transactions on Green Communications and Networking. In January 2016, he was elevated to Fellow of the IEEE for research contributions to antenna arrays and multi-hop communications. He wa also one of four scientists elevated to Fellow of the European Association for Signal Processing (EURASIP) in 2023 for ``Signal Processing Advances in Multiple Antenna and relayed Wireless Communication Systems". 
\end{IEEEbiography}

\begin{IEEEbiography}[{\includegraphics[width=1in,height=1.25in,clip,keepaspectratio]{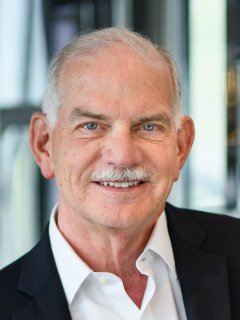}}]{H. Vincent Poor}
	(S’72, M’77, SM’82, F’87) received the Ph.D. degree in EECS from Princeton University in 1977.  From 1977 until 1990, he was on the faculty of the University of Illinois at Urbana-Champaign. Since 1990 he has been on the faculty at Princeton, where he is currently the Michael Henry Strater University Professor. During 2006 to 2016, he served as the dean of Princeton’s School of Engineering and Applied Science, and he has also held visiting appointments at several other universities, including most recently at Berkeley and Cambridge. His research interests are in the areas of information theory, machine learning and network science, and their applications in wireless networks, energy systems and related fields. Among his publications in these areas is the book \textit{Machine Learning and Wireless Communications}.  (Cambridge University Press, 2022). Dr. Poor is a member of the National Academy of Engineering and the National Academy of Sciences and is a foreign member of the Royal Society and other national and international academies. He received the IEEE Alexander Graham Bell Medal in 2017. 
\end{IEEEbiography}

\end{document}